\documentclass[11pt,a4paper]{article}
\pdfoutput=1

\usepackage{setup}
\usepackage{jheppub}

\subheader{\small IPPP/17/58, ZU-TH 21/17}

\title{\boldmath Precise predictions for the angular coefficients in $\PZ$-boson production at the LHC}

\author[a]{R. Gauld,}
\author[a,b]{A. Gehrmann--De Ridder,}
\author[b]{T.~Gehrmann,}
\author[c]{E.~W.~N.~Glover,}
\author[a]{A. Huss}

\affiliation[a]{Institute for Theoretical Physics, ETH, CH-8093 Z\"urich, Switzerland}
\affiliation[b]{Department of Physics, University of Z\"urich, CH-8057 Z\"urich, Switzerland}
\affiliation[c]{Institute for Particle Physics Phenomenology, Durham University,  Durham DH1 3LE, UK}

\emailAdd{rgauld@phys.ethz.ch}
\emailAdd{gehra@phys.ethz.ch}
\emailAdd{thomas.gehrmann@uzh.ch}
\emailAdd{e.w.n.glover@durham.ac.uk}
\emailAdd{ahuss@phys.ethz.ch}

\abstract{
The angular distributions of lepton pairs in the Drell--Yan process can provide rich information
on the underlying QCD production mechanisms. These dynamics can be parameterised in 
terms of a set of frame dependent angular coefficients, $A_{i=0,\ldots,7}$, which depend on the 
invariant mass, transverse momentum, and rapidity of the lepton pair.
Motivated by recent measurements of these coefficients by ATLAS and CMS, and in particular by the apparent 
violation of the Lam--Tung relation $A_0-A_2=0$, we perform a precision study of the angular coefficients at $\cO(\alphas^3)$ 
in perturbative QCD.
We make predictions relevant for $\Pp\Pp$ collisions at $\sqrt{s} = 8~\TeV$, and perform comparisons 
with the available ATLAS and CMS data as well as providing predictions for a prospective measurement at LHCb.
To expose the violation of the Lam--Tung relationship we propose a new observable $\Delta^\mathrm{LT} = 1-A_2/A_0$ that 
is more sensitive to the dynamics in the region where $A_0$ and $A_2$ are both small.
We find that the $\cO(\alphas^3)$ corrections have an important impact on the $\ptz$ distributions for 
several of the angular coefficients, and are essential to provide an adequate description of the data.
The compatibility of the available ATLAS and CMS data is reassessed by performing a partial $\chi^2$ test 
with respect to the central theoretical prediction which shows that $\chi^2/N_\mathrm{data}$ is significantly 
reduced by going from ${\cO}(\alphas^2)$ to ${\cO}(\alphas^3)$.
}


\newcommand{\yz}{\ensuremath{y_\PZ}\xspace}

\newcommand{\ptz}{\ensuremath{p_{\rT,\PZ}}\xspace}
\newcommand{\phistar}{\ensuremath{\phi_{\eta}^*}\xspace}
 
\newcommand{\tg}{\tilde{g}}
\newcommand{\tp}{\tilde{p}}

\begin{document}
\maketitle
\flushbottom


\section{Introduction} 
\label{sec1}

The production of $\PZ$ bosons followed by subsequent leptonic decay is a benchmark process at hadron colliders.
The production rate for this process is extremely
large, and, combined with the fact that the final state is
clean experimentally, it has allowed precise
(multi-) differential $\PZ$-boson cross section measurements 
to be performed both at the Tevatron~\cite{Aaltonen:2010zza,Abazov:2007jy} 
and the LHC~\cite{Aad:2014xaa,Khachatryan:2015oaa,Chatrchyan:2011wt,Aad:2015auj,Aaij:2015gna,Aaij:2015zlq,Aaij:2016mgv}.
Typically, these measurements are performed inclusively
with respect to the kinematic information of the gauge boson decay 
and have a wide range of phenomenological applications, including PDF and luminosity determinations.

Additional tests of the QCD dynamics for $\PZ$-boson production can also be performed by explicitly studying the angular distribution of the final-state leptons~\cite{Collins:1977iv,Lam:1978pu,Lam:1978zr,Lam:1980uc,Mirkes:1992hu,Mirkes:1994eb,Mirkes:1994dp}.
A prime example being the measurement of the forward--backward asymmetry
in lepton-pair production, differential in the lepton polar angle,
which provides important information on the coupling structure of the $\PZ$ boson to fermions~\cite{Chatrchyan:2011ya,
Aaltonen:2014loa,Abazov:2014jti,Aad:2015uau,Aaij:2015lka}.
However, an even richer structure is accessible by retaining the full 
differential information of the lepton kinematics. Under the assumption that the 
lepton pair is produced through the exchange of a gauge boson, the reconstructed 
lepton kinematics provide a direct probe of the polarisation of the intermediate gauge boson, 
which in turn exposes the underlying QCD production mechanism.
The QCD dynamics of this process can be expressed in terms of a set of eight frame dependent
angular coefficients $A_{i=0,\ldots,7}$, which depend on the 
invariant mass, transverse momentum, and rapidity of the lepton pair
and describe the production of the
intermediate gauge boson. 

The angular coefficients $A_0$ and $A_2$ further satisfy an important relation known 
as the Lam--Tung relation~\cite{Lam:1978pu,Lam:1978zr,Lam:1980uc}, $A_0-A_2=0$.
In the framework of perturbative QCD (pQCD), this relation can be shown to hold up to $\cO(\alphas)$ 
and is violated only at $\cO(\alphas^2)$ and higher.
At leading order, $A_0=A_2$ as a direct consequence of the spin-$\tfrac{1}{2}$ nature of the quarks,
and is further preserved at $\cO(\alphas)$ due to the vector-coupling of the spin-1 gluon to quarks.

Distributions for the angular coefficients can be extracted experimentally through 
fits to the measured final-state lepton kinematics, which can then be compared 
to the corresponding predictions obtained in pQCD.
The measurement of these angular coefficients is therefore interesting in its own right 
and much effort has been devoted to their precise determination.
Moreover, such a measurement also plays an important role 
in the determination of the $\PW$-boson mass $M_\PW$ at hadron colliders.
Indeed, a precise extraction of $M_\PW$ requires control of the Monte Carlo samples used to 
describe the kinematic distribution of leptons resulting from $\PW$-boson decay. The approach 
to generating these samples (and/or reweighting them) can in part be validated by using 
the  $\PZ$-boson production process as a case study, where the predicted values for all relevant angular coefficients 
can be directly compared to data.

On the theoretical side, the angular coefficients have been computed in pQCD up to $\cO(\alphas)$~\cite {Collins:1977iv,Lam:1978pu,Lam:1978zr,Lam:1980uc}
and $\cO(\alphas^2)$~\cite{Mirkes:1992hu,Mirkes:1994eb,Mirkes:1994dp} for non-vanishing transverse momenta $\ptz$ of the $\PZ$ boson.
For the inclusive Drell--Yan process, the $\cO(\alphas^2)$ corrections are available 
in the parton-level generators DYNNLO~\cite{Catani:2009sm} and FEWZ~\cite{Gavin:2010az},
which retain the full kinematical information of the final state and allow for a direct comparison to data in the fiducial region. 
These fixed-order predictions have been further matched to parton showers at NNLO in Ref.~\cite{Karlberg:2014qua}, 
where a comparison to the angular coefficients has also been performed.
Using the results obtained with DYNNLO and FEWZ, a detailed comparison 
to all available hadron collider and fixed target data has been carried out in Ref.~\cite{Lambertsen:2016wgj}. 
Studies of the Lam--Tung relation in the context of the intrinsic transverse momentum of the parton have 
also recently been considered in Ref.~\cite{Motyka:2016lta}.

Experimentally, a number of the angular coefficients were determined
in fixed target experiments by the NA10~\cite{Guanziroli:1987rp}, E615~\cite{Conway:1989fs},
and FNAL E866/NuSea~\cite{ Zhu:2006gx, Zhu:2008sj} collaborations 
using a variety of beams (pions, protons) and targets (tungsten, deuterium).
It is worth noting that the kinematical range probed in these fixed-target experiments was restricted to small invariant masses of the lepton pairs, 
typically from a few \GeV up to $\sim15~\GeV$.
In this regime, photon-exchange in the Drell--Yan process is by far the dominant contribution and
only the parity-even angular coefficients could be determined.

At high-energy colliders such as the Tevatron and the LHC, on the other hand, lepton-pair invariant masses around the $\PZ$-boson mass are considered, which are dominated by $\PZ$-boson exchange and also allow for the study of the parity-odd angular coefficients.
The measurement of angular coefficients at hadron colliders were performed 
by the CDF~\cite{Aaltonen:2011nr} collaboration in $\Pp\Pap$ collisions at a centre-of-mass (CoM)
energy of $\sqrt{s} = 1.96~\TeV$, and more recently by the CMS~\cite{Khachatryan:2015paa} and 
ATLAS~\cite{Aad:2016izn} collaborations in $\Pp\Pp$ 
collisions at $\sqrt{s} = 8~\TeV$.
Each of these analyses were performed in an invariant-mass window around the $\PZ$-boson resonance and in the Collins--Soper reference frame~\cite{Collins:1977iv}. 
Most notably, both ATLAS and CMS observe for the first time clear evidence for 
the violation of the Lam--Tung relation in $\PZ$-boson production.%
\footnote{
Note that this effect had been already observed  
by NA10~\cite{Guanziroli:1987rp} and E615~\cite{Conway:1989fs} for low-mass lepton pairs, whereas 
both FNAL E866/NuSea~\cite{Zhu:2006gx, Zhu:2008sj} and CDF~\cite{Aaltonen:2011nr} found results 
consistent with the difference $(A_0-A_2)$ being zero.
}
The new results from ATLAS and CMS are therefore particularly interesting as they 
demonstrate the violation of the Lam--Tung relation at energies never probed before.  

However, compared to the fixed-order $\cO(\alphas^2)$ prediction for 
lepton-pair production using the fixed order parton level code DYNNLO~\cite{Catani:2009sm}, a ``significant deviation'' is reported by the ATLAS collaboration~\cite{Aad:2016izn} for the difference $(A_0-A_2)$ in the region with $\ptz >20~\GeV$. 
Although less significant, a similar trend is also observed in the CMS data~\cite{Khachatryan:2015paa} where the $\cO(\alphas^2)$ prediction for the Drell--Yan pair production reaction is obtained 
using the parton-level generator FEWZ~\cite{Gavin:2010az}. 
Both experiments observe that the data exceeds the corresponding theory prediction for this observable. 
A tension is also observed in the \ptz spectrum for the angular coefficient $A_2$, where the data tends to undershoot the theory prediction.   
It is worth noting that although both FEWZ and DYNNLO yield predictions which are accurate at
next-to-next-to-leading order (NNLO) for the inclusive $\PZ$-boson production cross section, in analogy 
to the case for the \ptz or \phistar distributions studied in Refs.~\cite{Ridder:2016nkl,Gehrmann-DeRidder:2016jns}, 
these codes provide NLO-accurate predictions for the angular coefficients $A_i$ and LO-accurate 
predictions for the difference $(A_0 -A_2)$ (since $A_0 = A_2$ 
up to $\cO(\alphas)$ by virtue of the Lam--Tung relation).

The purpose of this work is to reassess the compatibility of the LHC data
to theory by providing predictions for the phenomenologically most important angular coefficients 
in high-mass lepton pair production at $\cO(\alphas^3)$, while focussing on the kinematic region with $\ptz >10~\GeV$, 
where many of the angular coefficients start to acquire non-vanishing values. 
This accuracy is achieved through the calculation of the $\PZ+\jet$ process at 
$\cO(\alphas^3)$~\cite{Ridder:2015dxa} at finite \ptz without requiring any resolved jets in the final state.

The layout of the paper is as follows. In Section~\ref{sec2}, the theoretical formalism for 
decomposing $\PZ$-boson production in terms of angular coefficients and spherical
harmonics is discussed. 
We further propose a new observable $\Delta^\mathrm{LT}$, which is particularly suited to 
study the violation of the Lam--Tung relation in the low-$\ptz$ regime.
Numerical predictions for these angular coefficients are provided
in Section~\ref{sec3}, along with a detailed comparison to the available LHC data. In addition, 
predictions for the LHCb experiment are provided, for which no measurement is available at present.
A summary of our findings and concluding remarks are presented in Section~\ref{sec:4}.


\section{Theoretical preliminaries}
\label{sec2}

\begin{figure}
  \begin{minipage}[t]{.48\linewidth}
  	\centering
  	\includegraphics[scale=1.2]{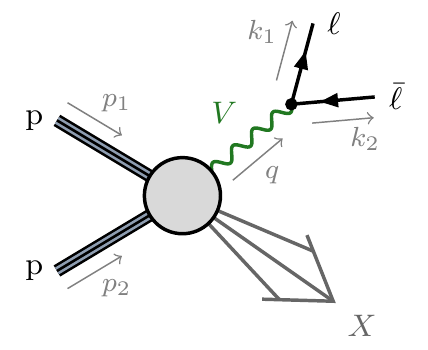}
  	\caption{Schematic diagram illustrating the kinematic configuration for the process.}
  	\label{fig:pp_VX}
  \end{minipage}
  \hfill
  \begin{minipage}[t]{.48\linewidth}
  	\centering
  	\includegraphics[width=\linewidth]{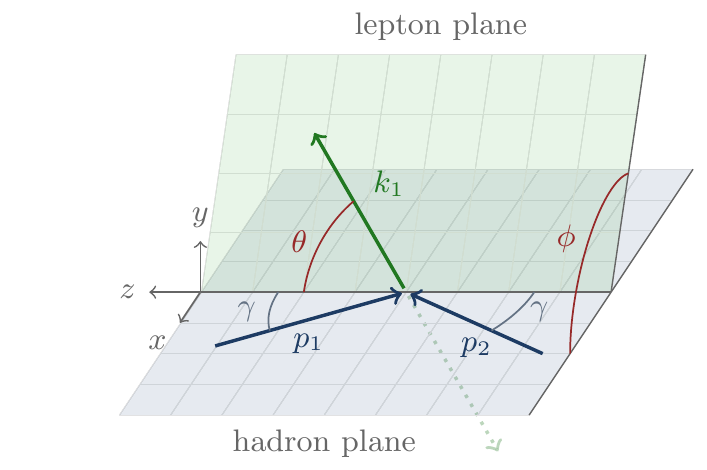}
  	\caption{The definition of the Collins--Soper~\cite{Collins:1977iv} angles in the di-lepton rest frame.}
  	\label{fig:CS_frame}
  \end{minipage}
\end{figure}

We consider the inclusive production of lepton pairs through the decay of an intermediate gauge boson, $\Pp(p_1) + \Pp(p_2) \to V(q) + X \to \Pl(k_1) + \Pal(k_2) + X$ as depicted in Fig.~\ref{fig:pp_VX}.
The cross section for this process can be written as the contraction of a lepton tensor ($L^{\mu\nu}$) describing the final-state decay with a hadronic tensor ($H_{\mu\nu}$) that describes the production sub-process, namely $L^{\mu\nu} \; H_{\mu\nu}$.
The lepton tensor in this context takes the role of an analyser, providing a probe of the structure of $H_{\mu\nu}$.
Note that the definition of the hadronic tensor includes the convolution with the PDFs as well as the integral over 
any degrees of freedom associated with the hadronic recoil ``$+X$''. 
As a result, $H_{\mu\nu}$ only depends on the four-momenta $p_1$, $p_2$, and $q$. 
Based on Lorentz- and gauge-invariance, the general decomposition of the hadronic tensor into form factors therefore reads%
\footnote{Owing to $H_{\mu\nu}^* = H_{\nu\mu}$, the symmetric and anti-symmetric parts of the hadronic tensor are purely real and imaginary, respectively.}
\begin{align}
  H_{\mu\nu} &=
  H_1 \; \tg_{\mu\nu} 
  + H_2 \; \tp_{1,\mu} \, \tp_{1,\nu} 
  + H_3 \; \tp_{2,\mu} \, \tp_{2,\nu} 
  + H_4 \; ( \tp_{1,\mu} \, \tp_{2,\nu} + \tp_{2,\mu} \, \tp_{1,\nu} ) \nonumber \\ &\quad 
  + \ri\, H_5 \; ( \tp_{1,\mu} \, \tp_{2,\nu} - \tp_{2,\mu} \, \tp_{1,\nu} ) 
  + \ri\, H_6 \; \epsilon(\mu, \nu, p_1, q) 
  + \ri\, H_7 \; \epsilon(\mu, \nu, p_2, q) \nonumber \\ &\quad 
  + H_8 \; \bigl(\, \tp_{1,\mu} \, \epsilon(\nu, p_1, p_2, q) + \mu \leftrightarrow \nu \,\bigr) 
  + H_9 \; \bigl(\, \tp_{2,\mu} \, \epsilon(\nu, p_1, p_2, q) + \mu \leftrightarrow \nu \,\bigr) ,
  \label{eq:Hi}
\end{align}
with $\tg_{\mu\nu} = g_{\mu\nu} - \frac{q_\mu q_\nu}{q^2}$ and $\tp_\mu = \tg_{\mu\nu} p^\nu$.
The decomposition~\eqref{eq:Hi} further incorporates discrete symmetries such that $H_{1,\ldots,5}$ ($H_{6,\ldots,9}$) and $H_{5,8,9}$ ($H_{1,\ldots,4,6,7}$) are respectively even (odd) under parity and time-reversal.

It is interesting to note that lepton-pair production satisfies an analogous relation to the Callan--Gross relation in deep-inelastic scattering~(DIS) known as the Lam--Tung relation~\cite{Lam:1978pu,Lam:1978zr,Lam:1980uc}, 
\begin{align}
  H_1 &= \frac{1}{2} \; H^{\mu}{}_{\mu} \,.
  \label{eq:LT_cov}
\end{align}
This relation, formulated in a covariant manner, is frame independent and characteristic of the spin-$\tfrac{1}{2}$ nature of the quark. 
It has been further shown~\cite{Lam:1980uc} that Eq.~\eqref{eq:LT_cov} is not affected by $\cO(\alphas)$ QCD corrections,%
\footnote{
In the DIS process, the Born kinematics are highly constrained and 
are necessarily part of the Callan--Gross relation. In the presence of
real-emission corrections, these constraints are lifted leading to
a violation of the Callan--Gross relation at $\cO(\alphas)$.
}
which follows as a direct consequence of the vector-coupling of the spin-1 gluon to quarks~\cite{ArteagaRomero:1983ji}.
However, this relation has been shown to be violated at $\order{\alphas^2}$~\cite{Mirkes:1994dp}.
As such, the Lam--Tung relation offers a unique opportunity to study the pQCD predictions of the underlying dynamics encoded in $H_{\mu\nu}$ in more detail than through rate measurements alone.

To further elucidate the Lam--Tung relation, let us consider the kinematics of this process in
the lepton-pair rest frame where the final-state lepton momenta can be expressed in terms of the angles $\theta$ and $\phi$:
\begin{align}
  k^\mu_{1,2} &= \frac{Q}{2} \; (1, \pm\sin\theta \cos\phi, \pm\sin\theta \sin\phi, \pm\cos\theta)^\rT , &
  Q &= \sqrt{q^2} ,
\end{align}
where so far the orientation of the coordinate axes remains unspecified.
The only non-vanishing entries of the hadronic tensor~\eqref{eq:Hi} are the space--space components $H_{ij}$ ($H_{\mu 0} = H_{0 \nu} = 0$) ,
\begin{equation}
  H_{\mu\nu} \;\xrightarrow{\vec{q}=0}\; 
  \begin{pmatrix}
  0 & 0 & 0 & 0 \\
  0 & H_{11} & H_{12} & H_{13} \\
  0 & H_{21} & H_{22} & H_{23} \\
  0 & H_{31} & H_{32} & H_{33} 
  \end{pmatrix} ,
\end{equation}
in this reference frame.
After contracting the hadronic tensor with the lepton tensor of the $\PZ\to\Plm\Plp$ decay, the cross section can be decomposed in terms of spherical harmonics of up to degree two according to
\begin{align} 
  \frac{ \rd\sigma}{\rd^4q~\rd\cos\theta~\rd\phi} &= 
  \frac{3}{16 \pi} \; \frac{\rd\sigma^\mathrm{unpol.}}{\rd^4q}  
  \; \bigg\{ 
  (1 + \cos^2\theta) 
  + \frac{1}{2}\ A_0 \ (1 - 3 \cos^2\theta)  \nonumber \\ & \quad
  + A_1 \ \sin (2 \theta) \cos\phi  
  + \frac{1}{2}\ A_2 \ \sin^2\theta\ \cos (2 \phi)  \nonumber \\ & \quad
  + A_3 \ \sin \theta\ \cos \phi 
  + A_4 \ \cos\theta  
  + A_5 \ \sin^2\theta \ \sin (2 \phi)  \nonumber \\ & \quad
  + A_6 \ \sin (2 \theta)\ \sin \phi 
  + A_7 \ \sin \theta \ \sin \phi 
  \bigg\} ,
  \label{eq:Ai}
\end{align}
where $\rd\sigma^\mathrm{unpol.}$ is the unpolarised cross section. 
We note that the first term inside the parenthesis equal to $(1 + \cos^2\theta)$ is not accompanied by a separate angular coefficient, as its normalisation is described by $\rd\sigma^\mathrm{unpol.}$ that has been extracted as a global pre-factor in Eq.~\eqref{eq:Ai}.
The unpolarised cross section is given by the trace of the hadronic tensor and for $\PZ$ exchange explicitly reads
\begin{align}
  \frac{\rd\sigma^\mathrm{unpol.}}{\rd^4q}
  &= \frac{32\pi^2}{3} \; \alpha \;\bigl[ (g_\Pl^+)^2+(g_\Pl^-)^2 \bigr]
  \; Q^2 \;
  (H_{11}+H_{22}+H_{33}) ,
  \label{eq:unpol}
\end{align}
with $\alpha$ denoting the fine-structure constant and $g_\Pl^{\pm}$ the chiral $\PZ$-boson couplings to charged leptons.%
\footnote{
  In general, when both $\PZ$- and $\Pgg$-exchange are considered, the total cross section in this formalism can be expressed as a sum of three terms.
  Each comprises independent contractions between lepton and hadronic tensors associated to the photon-exchange, $\PZ$-exchange, and the $\PZ$--$\Pgg$-interference.
}
These couplings are defined according to
\begin{align}
  g_\Pl^+ &= \frac{\sw}{\cw} , &
  g_\Pl^- &= \frac{\sw^2-\frac{1}{2}}{\sw\cw} , &
  \sw &\equiv \sin\theta_\rw, \quad
  \cw \equiv \cos\theta_\rw ,
\end{align}
where $\theta_\rw$ is the weak mixing angle.

The form factors $H_{1,\ldots,9}$ in Eq.~\eqref{eq:Hi}, or equivalently the nine non-vanishing components $H_{ij}$ of the hadronic tensor, are directly related to the eight angular coefficients $A_{0,\ldots,7}$ and the unpolarised cross section.
Explicitly, the $A_i$ are given by
\begin{align}
  A_0 &= 2 \; H_{33}            \,\;c_{+}, &
  A_1 &= - (H_{13}+H_{31})      \,\;c_{+}, & 
  A_2 &= 2 (H_{22}-H_{11})      \,\;c_{+}, \nonumber \\ 
  A_3 &= 2 \ri (H_{23}-H_{32})  \,\;c_{-}, & 
  A_4 &= 2 \ri (H_{12}-H_{21})  \,\;c_{-}, & 
  A_5 &= - (H_{12}+H_{21})      \,\;c_{+}, \nonumber \\ 
  A_6 &= - (H_{23}+H_{32})      \,\;c_{+}, & 
  A_7 &= 2 \ri (H_{31}-H_{13})  \,\;c_{-},
  \label{eq:Ai_Hij}
\end{align}
where the proportionality factors $c_{\pm}$ arise from the fact that $\rd\sigma^\mathrm{unpol.}$ has been removed as a prefactor in the definition~\eqref{eq:Ai} and are given by
\begin{align}
  c_+ &= (H_{11}+H_{22}+H_{33})^{-1} , &
  c_- &= \frac{(g_\Pl^+)^2-(g_\Pl^-)^2}{(g_\Pl^+)^2+(g_\Pl^-)^2} \; (H_{11}+H_{22}+H_{33})^{-1} .
\end{align}
The unpolarised cross section~\eqref{eq:unpol} completes Eq.~\eqref{eq:Ai_Hij} as the ninth linearly independent combination of the $H_{ij}$.

Let us now choose a specific reference frame by defining the direction of the axes in the lepton-pair rest frame. 
To this end, we consider the Collins--Soper frame~\cite{Collins:1977iv} shown in Fig.~\ref{fig:CS_frame}:
The $z$-axis is chosen as the external bisector of the incoming beam directions, $\hat{e}_{z}^\mathrm{CS} \sim \pm (\vec{p}_1 - \vec{p}_2)$, where the positive $z$-direction is aligned with the $z$-direction of the lepton pair in the laboratory frame.
The $x$-axis lies in the hadron plane orthogonal to the $z$-axis and points in the direction of $\hat{e}_{x}^\mathrm{CS} \sim -(\vec{p}_1+\vec{p}_2)$. 
Lastly, the $y$-axis is chosen to complete a right-handed Cartesian coordinate system and is orthogonal to the hadronic event plane.
The four-momenta of the incoming hadrons in this reference frame are given by%
\footnote{
Note that we have suppressed the additional sign ambiguity in the $z$-component of $p_{1,2}^\mu$ 
due to the alignment of the $z$-axis w.r.t.\ the $\PZ$-boson direction in the laboratory frame.
} 

\begin{align}
  p_{1,2}^\mu &= E_{1,2} \; (1, -\sin\gamma, 0, \pm\cos\gamma)^\rT , & 
  E_{1,2} &= \frac{(q \cdot p_{1,2})}{Q} , &
  \cos\gamma &= \frac{Q}{\sqrt{Q^2 + q_\rT^2}} .
\end{align}

Returning to the Lam--Tung relation~\eqref{eq:LT_cov}, one can derive the corresponding relation in terms of the angular coefficients $A_i$ in the Collins--Soper frame
\begin{align}
  0 &\equiv
  2 H_1 - H^{\mu}{}_{\mu}  \nonumber \\
  &= 2 H_1 - H_1\, \tg^{\mu}{}_{\mu} - H_2\, \tp_1^2 - H_3\, \tp_2^2 -H_4\, 2 (\tp_1\cdot\tp_2)  \nonumber \\
  &= - H_1 +(E_1)^2 H_2 + (E_2)^2 H_3 + 2 E_1 E_2 \left( \sin^2\gamma - \cos^2\gamma \right) H_4  \nonumber \\
  &= H_{33} - H_{22} + H_{11}  \nonumber \\
  &\propto A_0 - A_2  ,
  \label{eq:LT_Ai}
\end{align}
where we have used
\begin{align}
  H_{11} &= -H_1 + \left[ (E_1)^2 H_2 + (E_2)^2 H_3 + 2 E_1 E_2 \,H_4 \right] \; \sin^2\gamma , \nonumber \\
  H_{22} &= -H_1 , \nonumber \\
  H_{33} &= -H_1 + \left[ (E_1)^2 H_2 + (E_2)^2 H_3 - 2 E_1 E_2 \,H_4 \right] \; \cos^2\gamma , 
  \label{eq:Hii}
\end{align}
for the non-vanishing diagonal components of the hadronic tensor.
We observe that the Lam--Tung relation is equivalent to $A_0 - A_2 = 0$. 
Note that the result of Eq.~\eqref{eq:LT_Ai} is not frame independent but only holds if both the $z$- and $x$-axis 
in the lepton-pair rest frame lie in the hadronic event plane.
This condition enters in the step where the form factors $H_i$ are expressed in terms of the diagonal $H_{ij}$ components using Eq.~\eqref{eq:Hii} and can be understood by inspecting the covariant formulation of Eq.~\eqref{eq:LT_cov} in the lepton-pair rest frame:
The only form factors that contribute to the trace of the hadronic tensor on the r.h.s.\ of Eq.~\eqref{eq:LT_cov} are $H_{1,\ldots,4}$.
The tensor structures multiplying $H_{2,3,4}$ only involve momenta lying inside the hadronic plane and it is solely the tensor $\tg_{\mu\nu}$ multiplying the form factor $H_1$ which has a non-vanishing component orthogonal to it.
The Lam--Tung relation therefore distinguishes the direction perpendicular to the hadronic plane and can be interpreted as a statement about the current--current correlation of the hadronic tensor in this direction.%
\footnote{For hypothetical spin-$0$ partons, the current correlator would be completely confined within the hadronic event plane, which then yields for Eq.~\eqref{eq:LT_cov}: $H_1=0$.}

Making use of the completeness of the spherical harmonics, the angular coefficients appearing in the decomposition provided
in Eq.~\eqref{eq:Ai} can be extracted through the projectors
\begin{align}
  A_0 &= 4-10 \, \avg{\cos^2\theta} , &
  A_1 &= 5 \, \avg{\sin(2\theta) \, \cos\phi} , &
  A_2 &= 10 \, \avg{\sin^2\theta \, \cos(2\phi)} , \nonumber \\
  A_3 &= 4 \, \avg{\sin\theta \, \cos\phi} , &
  A_4 &= 4 \, \avg{\cos\theta} , &
  A_5 &= 5 \, \avg{\sin^2\theta \, \sin(2\phi)} , \nonumber \\
  A_6 &= 5 \, \avg{\sin(2\theta) \, \sin\phi} , &
  A_7 &= 4 \, \avg{\sin\theta \, \sin\phi} , 
  \label{eq:Ai_proj}
\end{align}
where $\avg{\ldots}$ denotes taking the (normalised) weighted average over the angular variables $\theta$, $\phi$ and is defined as
\begin{align}
  \avg{f(\theta, \phi)} &\equiv 
  \frac{
    \int_{-1}^{1} \rd\cos\theta \int_0^{2\pi} \rd\phi \; 
    \rd\sigma(\theta,\phi) \;
    f(\theta, \phi)
  }{
    \int_{-1}^{1} \rd\cos\theta \int_0^{2\pi} \rd\phi \; 
    \rd\sigma(\theta,\phi)
  } .
  \label{eq:proj}
\end{align}

The dominant angular coefficients are $A_{0,\ldots,4}$, while $A_{5,6,7}$ vanish at $\order{\alphas}$ and only receive small $\order{\alphas^2}$ corrections from the absorptive parts of the one-loop amplitudes in $\PZ+\jet$ production.
We therefore will not discuss the coefficients $A_{5,6,7}$ in the following.
In the case of pure $\Pgg^*$ exchange, the relevant coefficients are the parity-conserving coefficients $A_{0,1,2}$.
$A_3$ and $A_4$, on the other hand, are odd under parity and proportional to the product of vector- and axial-vector-couplings of the gauge boson to the fermions.
As such, they are sensitive to the relative rate of incoming down- and up-type quark fluxes as well as the weak mixing angle $\sw$.
All the coefficients $A_i$ vanish in the limit $\ptz\to0$ with the exception of $A_4$, which is finite in this limit and directly related to the forward--backward asymmetry.

One of the goals of this work is to assess the compatibility of the observed extent of the Lam--Tung violation
with that expected in predictions based on pQCD. This can be done by directly studying the \ptz distribution for 
the difference of the angular coefficients $A_0$ and $A_2$. Here, we propose a new observable
\begin{align} \label{eq:dLT}
  \Delta^{\rm LT} \equiv 1 - \frac{A_2}{A_0} ,
\end{align}
which has the benefit that the strong suppression of the individual angular coefficients
in the low-$\ptz$ region is absent.
In addition, the dependence on the unpolarised cross section appearing in the denominator of Eq.~\eqref{eq:proj} cancels in the ratio between the two coefficients.
Consequently, this observable may help to expose the extent of the Lam--Tung
violation in this region.
In Section~\ref{sec3}, we shall compare our predictions to the available ATLAS and CMS data for the
$\ptz$ distributions of both $(A_0-A_2)$ and $\Delta^\mathrm{LT}$. In the latter case, 
the data will be re-expressed in terms of $\Delta^\mathrm{LT}$.


\section{Numerical predictions} 
\label{sec3}

In this section, we provide a comparison of the predictions for a set of angular coefficients
to the available ATLAS and CMS data in $\Pp\Pp$ collisions at $\sqrt{s} = 8~\TeV$. 
While the LHCb collaboration has not yet performed a measurement of the
angular coefficients, previous measurements of the $\PZ$-boson $p_\rT$ 
spectrum~\cite{Aaij:2015gna,Aaij:2015zlq,Aaij:2016mgv}
and forward--backward asymmetry~\cite{Aaij:2015lka} indicate that there is potential for such
a measurement in the forward region. We therefore also provide predictions
in the LHCb fiducial region at $\sqrt{s} = 8$~TeV. In all cases (ATLAS, CMS, LHCb), the angular coefficients 
are defined in the Collins--Soper reference frame~\cite{Collins:1977iv}.
In addition to the angular coefficients, we also provide absolute predictions for the unpolarised $\ptz$ distributions.%
\footnote{As compared to the results shown in Ref.~\cite{Ridder:2016nkl}, the kinematic setup differs slightly 
both for the ATLAS and CMS measurements and the theory uncertainty includes a seven-point scale variation.}
Special attention is also given to the difference $(A_0-A_2)$, 
where the quality of theoretical description with respect to the observed distributions is 
quantified by means of a $\chi^2$ test.
Furthermore, we also present a comparison to data for the new observable $\Delta^{\rm LT}$.

The measurements of the angular coefficients are performed differentially in \ptz and for various rapidity 
intervals, where in all cases an invariant-mass window for the lepton-pair final state is imposed around 
the $\PZ$-boson resonance. 
For non-vanishing values of \ptz, the LO prediction for this distribution can be obtained
from the $\cO(\alphas)$ tree-level $\PZ+\jet$ process, where the transverse 
momentum of the $\PZ$ boson is balanced with that of a single final-state QCD parton. 
The NLO QCD and EW corrections to this process have been computed in Refs.~\cite{Giele:1993dj,Denner:2011vu}, and
more recently the NNLO QCD corrections to this process have been completed~\cite{Ridder:2015dxa,Boughezal:2015ded}.
In this work, we employ the calculation of Ref.~\cite{Ridder:2015dxa} based on the antenna
subtraction formalism~\cite{GehrmannDeRidder:2005cm,GehrmannDeRidder:2005aw,GehrmannDeRidder:2005hi,Daleo:2006xa,Daleo:2009yj,Boughezal:2010mc,Gehrmann:2011wi,GehrmannDeRidder:2012ja,Currie:2013vh}
to provide NNLO-accurate QCD predictions for the \ptz distributions of the angular coefficients.
This process is implemented in the flexible parton-level Monte Carlo generator \textsc{NNLOjet}.

The predictions are provided in the $G_\mu$-scheme, where we take the following
set of numerical inputs: $M_{\PZ}^\mathrm{os} = 91.1876~\GeV$, $\Gamma_{\PZ}^\mathrm{os} = 2.4952~\GeV$, 
$M_{\PW}^\mathrm{os} = 80.385~\GeV$,  $\Gamma_{\PW}^\mathrm{os} = 2.085~\GeV,$ 
and $G_\mu = 1.16638 \cdot 10^{-5}~\GeV^{-2}$. In the extraction of the corresponding numerical values 
for $\alpha$ and $\sw^2$, 
we additionally include the dominant one- and two-loop universal corrections to the $\rho$-parameter~\cite{Fleischer:1993ub}
which relate $M_{\PW}-M_{\PZ}$ interdependence present beyond tree-level. Including these 
contributions leads to the effective values of $\alpha_\mathrm{eff.} = 0.007779$, $s_{\mathrm{w,eff.}}^2 = 0.2293$.

As a baseline PDF set, we use the central member of \verb|PDF4LHC15_nnlo_30|~\cite{Butterworth:2015oua,
Dulat:2015mca,Harland-Lang:2014zoa,Ball:2014uwa,Gao:2013bia,Carrazza:2015aoa}, and extract $\alphas$ from the grid provided with the PDF set---corresponding to $\alphas(Q=M_{\PZ}^\mathrm{os}) = 0.118$.

As discussed in Section~\ref{sec2}, the theoretical predictions for the coefficients $A_i$ can be obtained by computing the normalised expectation values of the spherical harmonics according to Eq.~\eqref{eq:proj}. To assess the theoretical uncertainty in the extraction of the coefficients through this method, various solutions are possible. In this paper, for comparison with LHC data, we choose to perform an independent variation of factorisation ($\muf$) and renormalization ($\mur$) scales in both the numerator and denominator of this expression.
The scales $\mur^{\rm{num.}}, \muf^{\rm{num.}}, \mur^{\rm{den.}},$ and $\muf^{\rm{den.}}$ are each independently varied by factors of $\tfrac{1}{2}$ and $2$ about the transverse energy of the lepton pair,
\begin{equation}
  \mu_0 \equiv E_{\rT,\PZ} = \sqrt{m_{\Pl\Pl}^2 + p_{\rT,\Pl\Pl}^2} \,
\end{equation}
with the constraint that all pairs of these \textbf{uncorrelated} scales satisfy $\tfrac{1}{2} \leq \mu^{i}_{a}/\mu^{j}_{b} \leq 2$.
In total this corresponds to 31 possible combinations and the associated uncertainty is obtained as 
the envelope around the central scale $\mur^{\rm{num.}} = \muf^{\rm{num.}} = \mur^{\rm{den.}} = \muf^{\rm{den.}}  = \mu_0$.

An alternative approach is to \textbf{correlate} the scale uncertainties between numerator and denominator. However, this
treatment can lead to an underestimation of the uncertainty due to missing higher-order effects. For example, 
at LO the renormalisation scale dependence is fully encapsulated in the strong coupling $\alpha_s(\mu)$ which entirely cancels
if the scales between numerator and denominator are \textbf{correlated}.

\begin{figure}
\centering
\includegraphics[width=.49\linewidth]{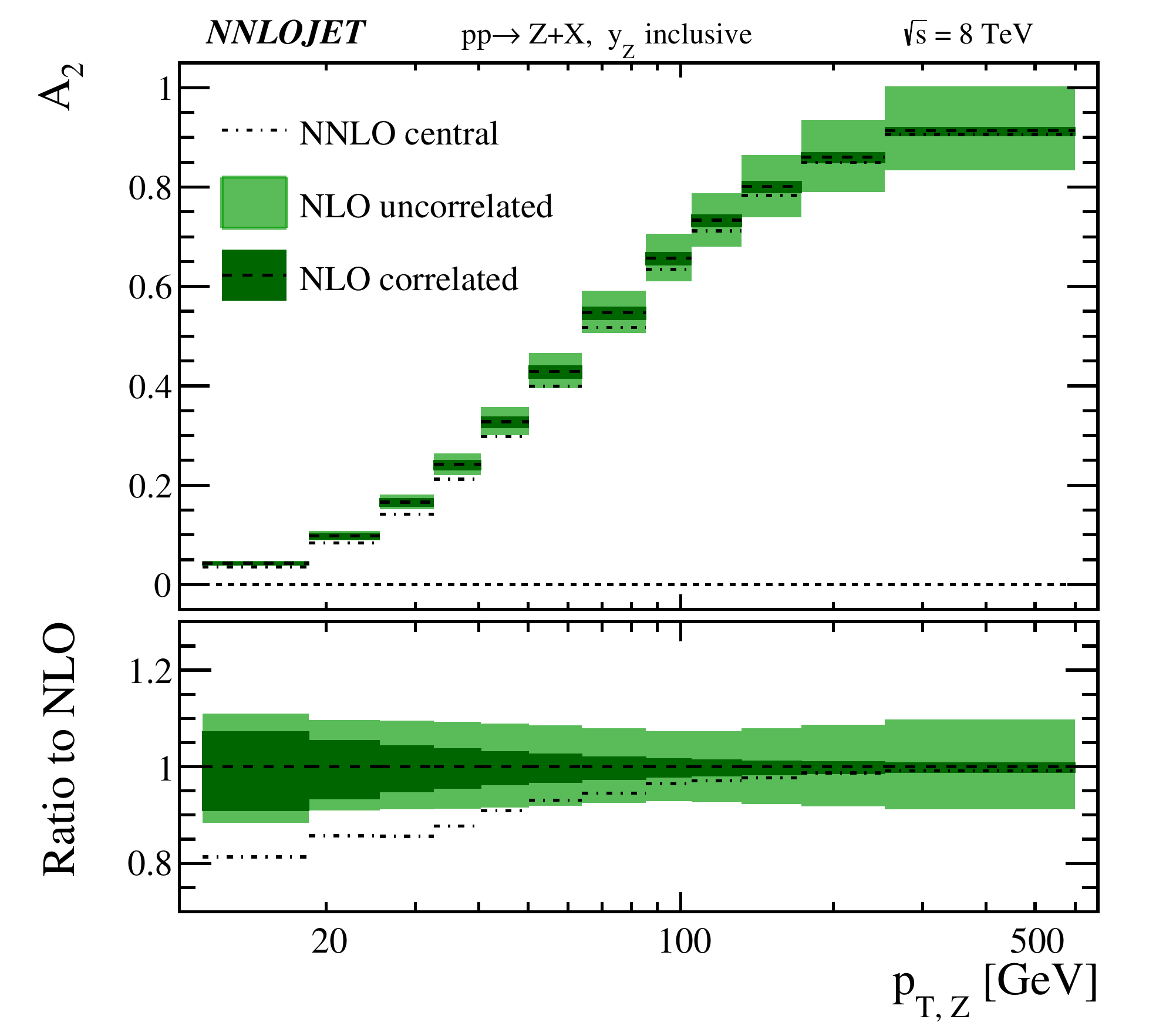} \hfill
\includegraphics[width=.49\linewidth]{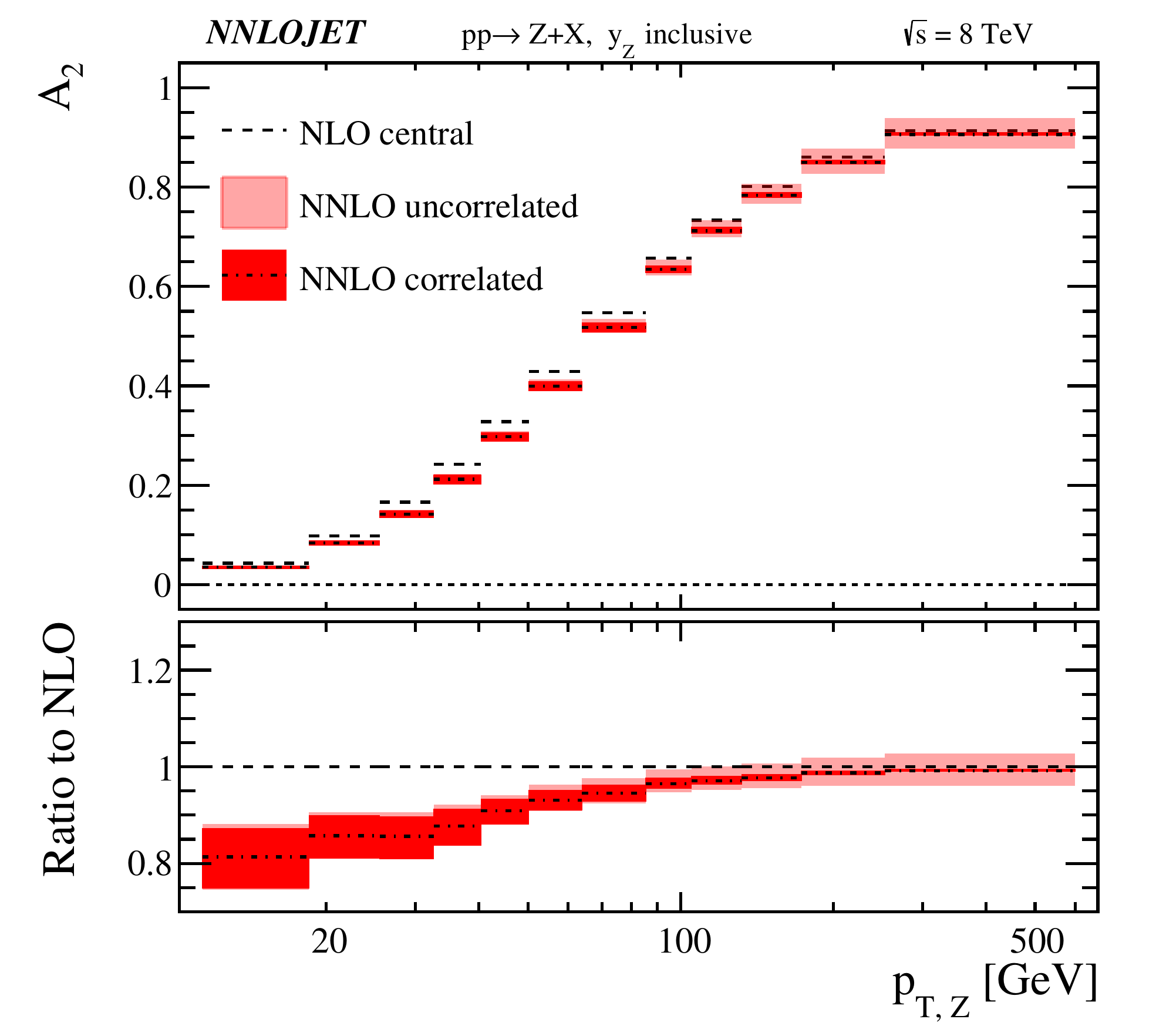} 
\caption{The \ptz distribution for the angular coefficient $A_2$ in $\Pp\Pp$ collisions at $\sqrt{s} = 8~\TeV$.
The uncertainty obtained when choosing to (un)correlate the scale choices in the extraction of $A_2$ is shown at 
NLO (left) and NNLO (right). In the lower panel, each distribution is shown normalised with respect
to the central NLO prediction.
}
\label{fig:A2_Scales}
\end{figure}

To further demonstrate this point, we show the impact of these two different approaches in Fig.~\ref{fig:A2_Scales} where the \ptz distribution for $A_2$ is evaluated at NLO and NNLO. The distributions are obtained with an invariant-mass cut of $80 < m_{\Pl\Pl} < 100~\GeV$ on the lepton-pair final state and inclusively with respect to rapidity of the lepton pair.
It is clearly seen that at NLO, these two prescriptions result in substantial differences in the scale uncertainty bands, with the correlated approach yielding considerably smaller uncertainty bands for \ptz above $20 ~\GeV$. At NNLO however, the scale uncertainty estimate of the $A_i$ coefficients obtained with either choice gives similar results in the low-pt region ($\ptz < 80$~GeV) where the NNLO effects are largest. Throughout this work, all distributions are obtained with the \textbf{uncorrelated} prescription discussed above.

It is worth commenting that the angular coefficients are evaluated differentially in \ptz
and in multiple kinematic regions, corresponding to the various experimental setups. 
In addition, each of these coefficients are computed through the projectors in Eq.~\eqref{eq:Ai_proj}, 
which are highly oscillating functions with respect to the leptonic kinematics, and consequently 
their stable numerical evaluation is rather challenging. This is particularly true for 
the difference $(A_0-A_2)$ for which large non-local cancellations
occur at the level of the angular coefficient as opposed to the integrand.

\subsection{Comparison to ATLAS data}
The ATLAS measurements have been performed with an invariant-mass cut of $80 < m_{\Pl\Pl} < 100~\GeV$
on the lepton-pair final state, and distributions for the angular coefficients in \ptz were extracted
for a range of different accessible rapidity regions. The data has also been presented integrated in \yz, 
obtained after performing an extrapolation to the full phase-space region. The comparison of the 
theoretical predictions is performed with respect to this data, referred to as \yz inclusive, for which 
the measurements are most precise.
The predictions of the various \ptz distributions are provided for the kinematic range of 
$\ptz \in [11.4,\,600]~\GeV$, and compared to the available data in this region. Furthermore, in the 
region of \ptz $< 85.4~\GeV$ we provide our predictions with a coarser choice of \ptz bins
with respect to the data, which are obtained by pair-wise combining neighbouring bins.
Before continuing, it is also important to highlight that we perform the comparison to the 
ATLAS data which is obtained prior to the regularisation procedure outlined as part of the experimental 
analysis---more detail can be found in Appendix~C of~\cite{Aad:2016izn}.
Our motivations for doing so are as follows. Firstly, the regularisation procedure introduces large
bin-to-bin correlations for the distributions of $A_i$ coefficients meaning that a visual comparison
to the regularised data can be misleading as large correlations are hidden from view. Secondly,
we wish to quantify the agreement between theory and data by performing a $\chi^2$ test, which
requires knowledge of the bin-to-bin correlations between the different $A_i$ coefficients in \ptz (this is 
particularly important if these correlations are large, which is the case for the regularised data). However, to our knowledge, 
a well-defined covariance matrix for the regularised version of the $A_i$ coefficients is not available.

\begin{figure}
\centering
\includegraphics[width=.49\linewidth]{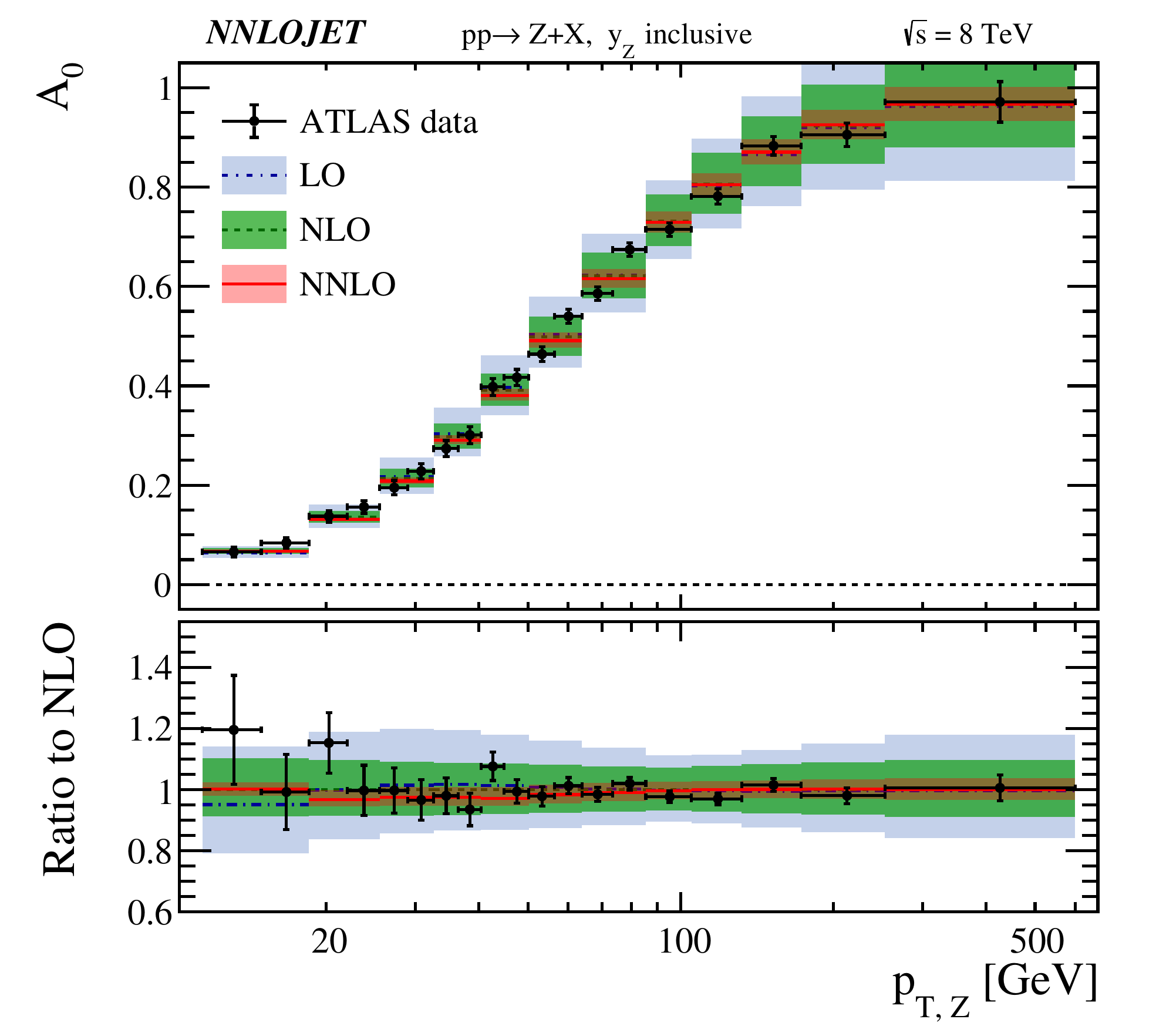} \hfill
\includegraphics[width=.49\linewidth]{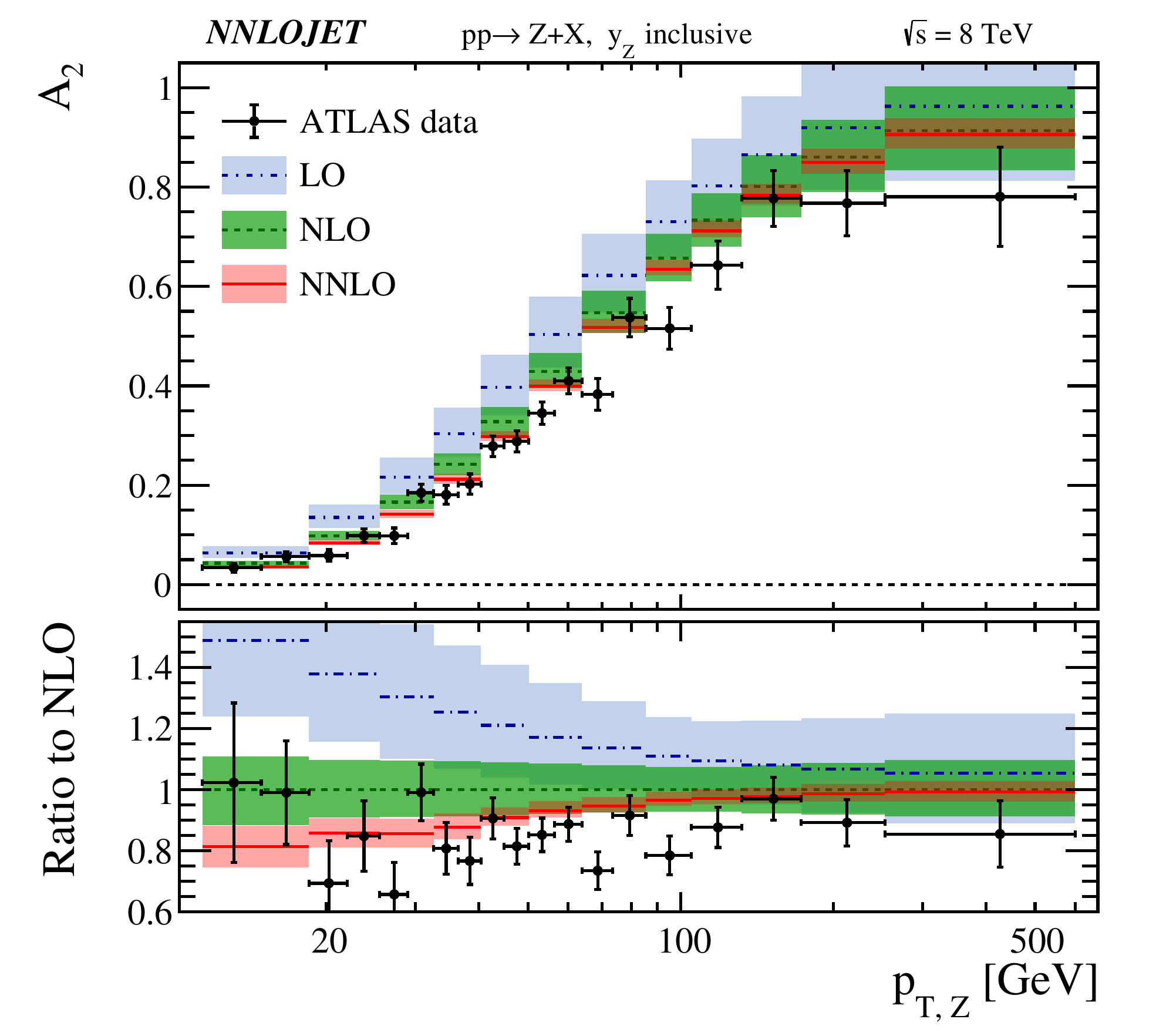} \\
\includegraphics[width=.49\linewidth]{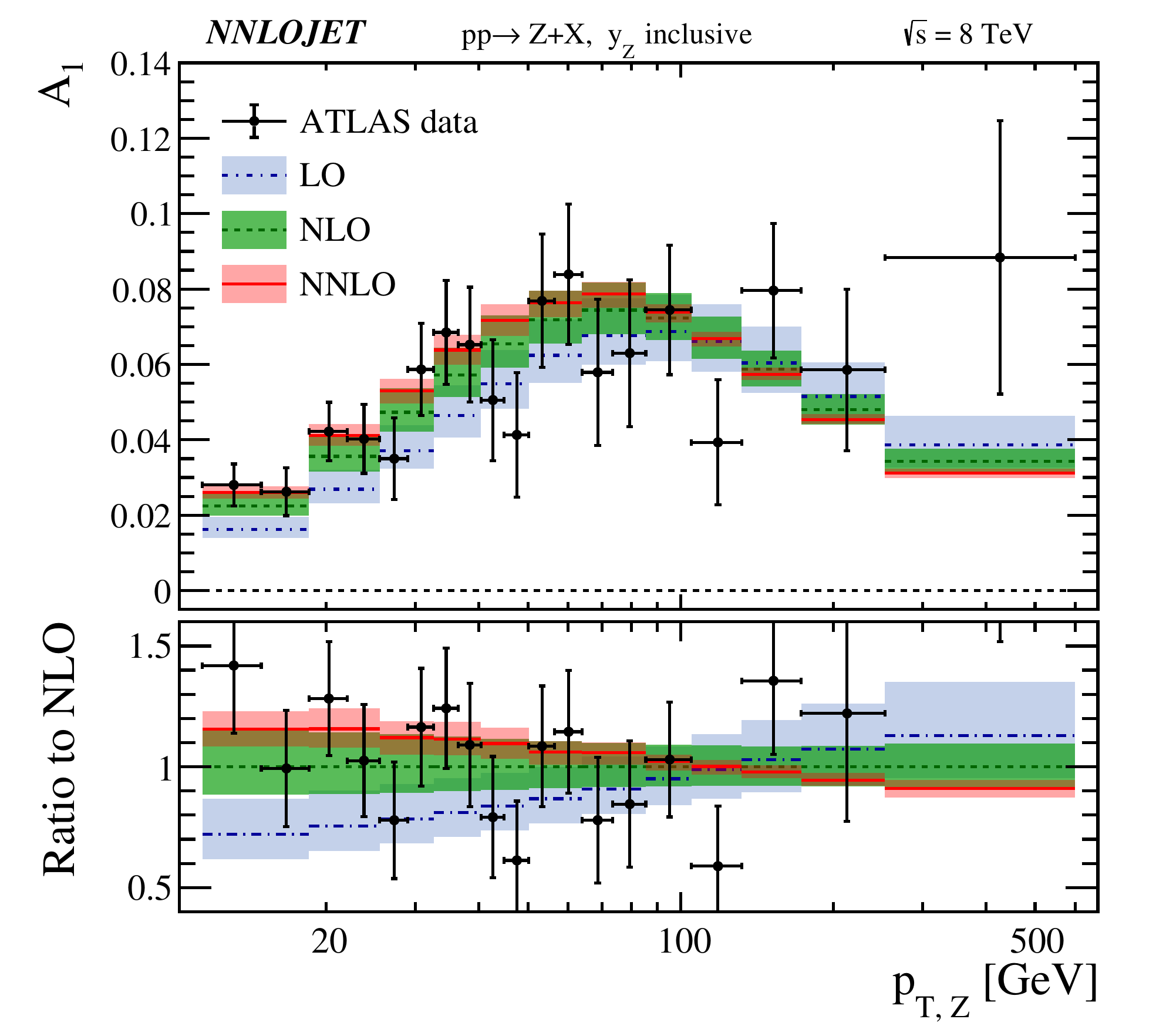} \hfill
\includegraphics[width=.49\linewidth]{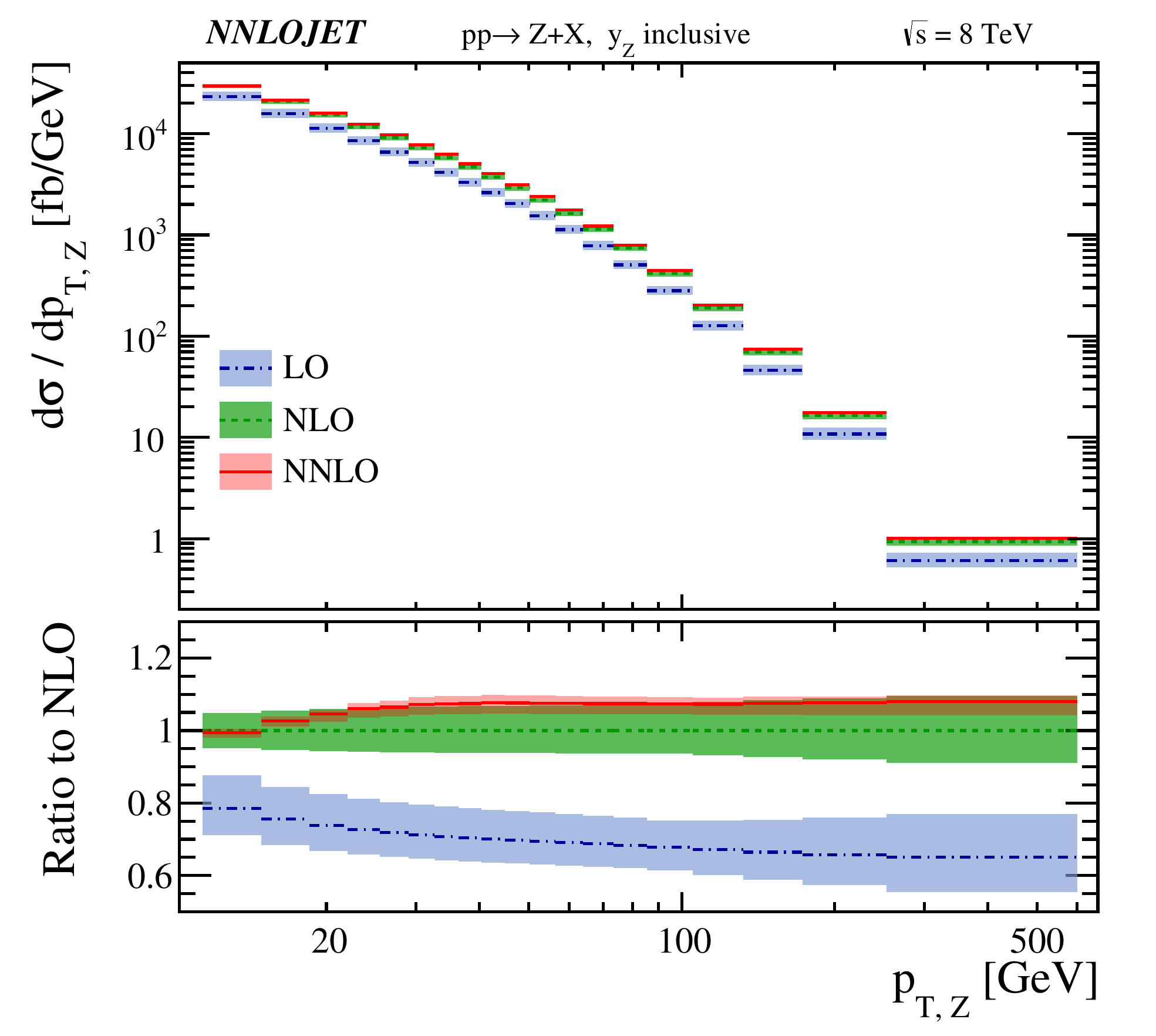} 
\caption{The \ptz distribution for the angular coefficients $A_0$ (upper left), $A_2$ (upper right), 
 $A_1$ (lower left), and the unpolarised cross section (lower right)  in $\Pp\Pp$ collisions at $\sqrt{s} = 8~\TeV$. 
 The ATLAS data (black points) are compared to the LO (blue fill), NLO (green fill), and NNLO (red fill)
 theoretical predictions. In the lower panel, each distribution is shown normalised 
 with respect to the central NLO prediction.
}
\label{fig:ATLAS_1}
\end{figure}
\begin{figure}
\centering
\includegraphics[width=.49\linewidth]{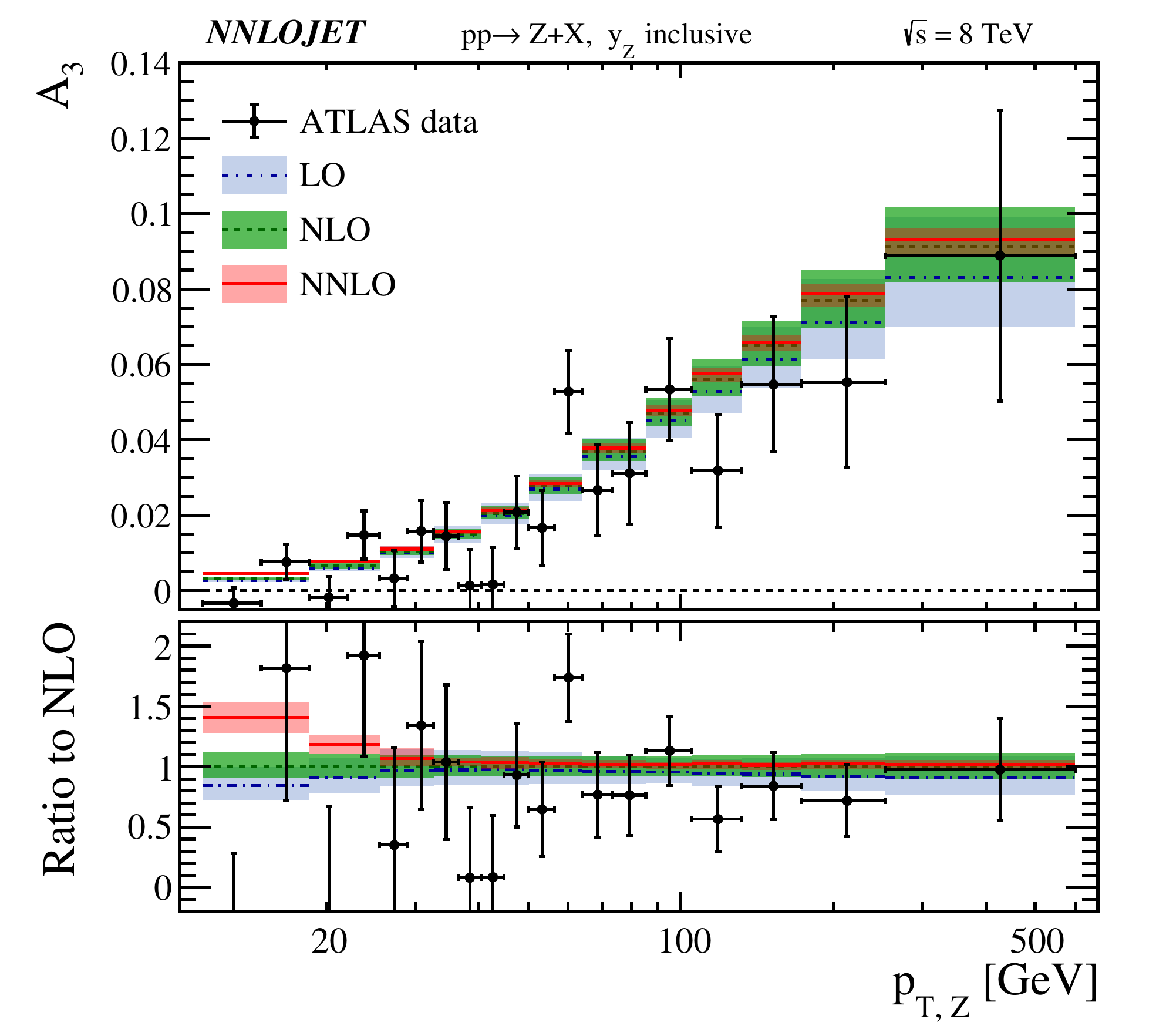} \hfill
\includegraphics[width=.49\linewidth]{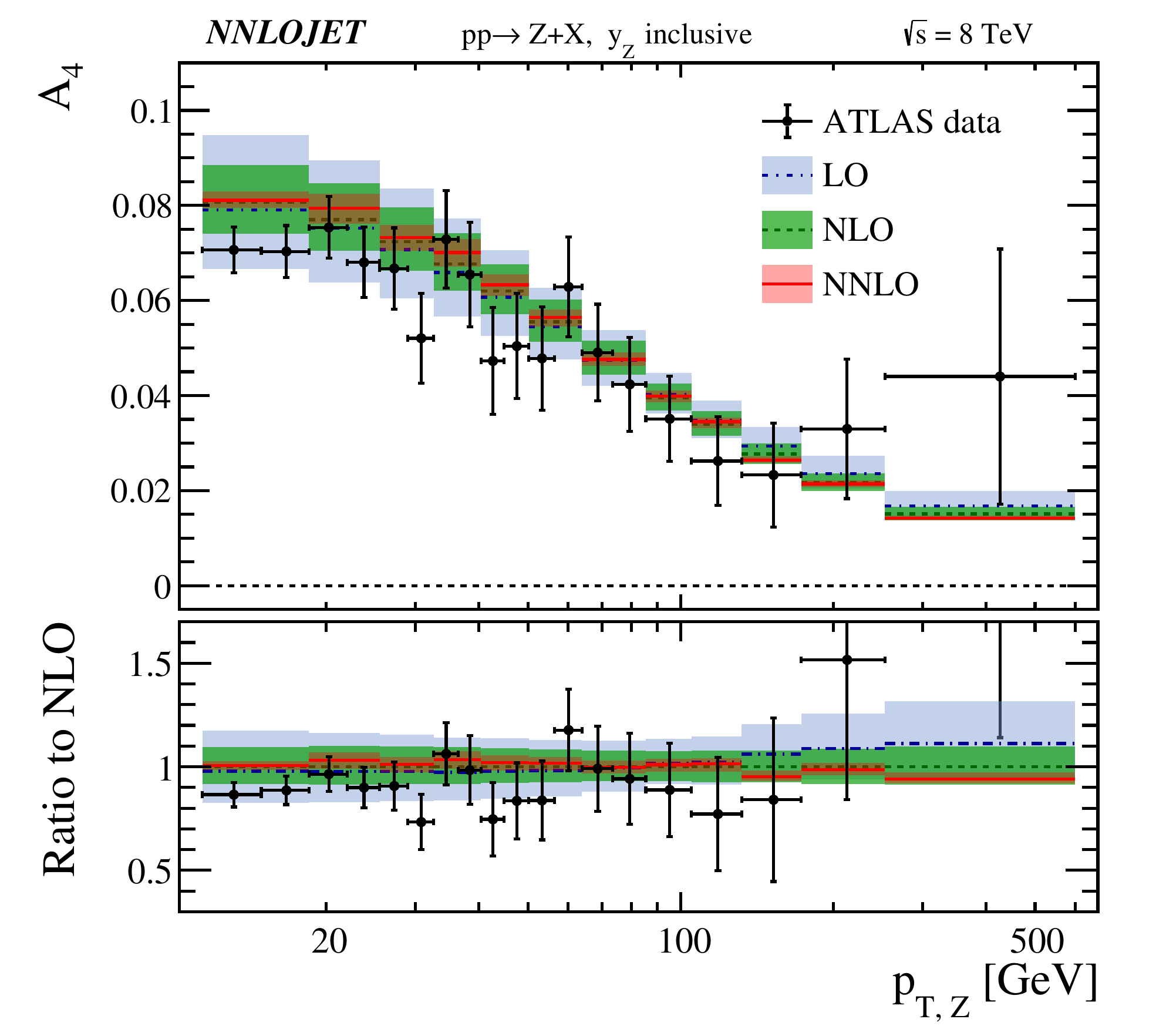}
\caption{
The \ptz distribution for the angular coefficients $A_3$ (left) and $A_4$ (right) 
in $\Pp\Pp$ collisions at $\sqrt{s} = 8~\TeV$. The ATLAS data (black points) are 
compared to the LO (blue fill), NLO (green fill), and NNLO (red fill) 
theoretical predictions. In the lower panel, each distribution is shown normalised 
with respect to the central NLO prediction.
}
\label{fig:ATLAS_2}
\end{figure}

The \ptz distributions for the angular coefficients $A_0$ (upper left), $A_1$ (lower left), 
$A_2$ (upper right), and the unpolarised cross section (lower right) are shown in Fig.~\ref{fig:ATLAS_1}. 
The ATLAS data is represented by black points, and is compared to theoretical predictions at LO (blue), 
NLO (green), and NNLO (red). In the lower panel of each plot, the same distributions are shown normalised
with respect to the central NLO prediction.

The NNLO corrections are observed to have an important impact on each distribution, substantially reducing the scale uncertainties in all cases. With respect to the central value at NLO, the NNLO corrections to $A_0$ are negative and typically below $5\%$ in magnitude. In the case of the $A_2$ distribution, the corrections are also negative and most sizeable in the region of \ptz $\in[10,\,50]$~GeV. The description of the observed $A_2$ distribution is visibly improved at NNLO, while the NLO predictions systematically overestimate the data.
It should be noted that the $y$-axis ranges for both $A_0$ and $A_2$ distributions are fixed to the same values to
allow a straightforward visual comparison of the relative impact of the NNLO corrections in each case
(this is also the reason why they are placed in neighbouring positions within the Figures).
In the case of the $A_1$ distribution, the corrections are positive at low $p_\rT$ and 
change sign to become negative at large $p_\rT$, resulting in a modified shape of the distribution.
The size of the corrections across the whole $\ptz$ range vary between $+10\%$ at low values of $\ptz$ and $-5\%$ 
in the last $\ptz$ bin shown.

In Figure~\ref{fig:ATLAS_2}, the same comparison is performed for the 
parity-violating angular coefficients $A_3$ (left) and $A_4$ (right).
The NNLO corrections reduce the scale uncertainty of the prediction,
while having little impact on the central value. With respect to the experimental precision, the NNLO corrections to these distributions (which are well described by the central NLO prediction) are phenomenologically unimportant for a comparison to data.
As discussed in Section~\ref{sec2}, these particular coefficients are sensitive to the product of vector- and axial-vector-couplings 
of the $\PZ$-boson to the initial-state quarks. The corresponding predictions for these distributions are 
therefore sensitive to a combination of the input value of $\sw^2$ as well as to the relative 
contribution of up- and down-type quark initiated processes.
To the accuracy of the experimental distributions for these coefficients, our
choice of input parameter scheme (including universal corrections via the $\rho$-parameter) 
provides a consistent description of the data. However, if the precision of future measurements of these 
coefficients improves, it would be important to revisit this comparison while including possibly 
also the effect of electroweak corrections and to assess the impact of PDF uncertainties on these distributions.
We note that while a measurement of these coefficients is sensitive to the weak mixing angle, 
a more direct extraction of this parameter is possible through the measurement
of the forward--backward asymmetry in lepton-pair production. Indeed, such a measurement 
has already been performed by the ATLAS collaboration~\cite{Aad:2015uau}.

\subsection{Comparison to CMS data}
A similar measurement of the angular coefficients has also been 
presented by the CMS collaboration~\cite{Khachatryan:2015paa}.
In this case, the angular coefficients have been measured differentially within rapidity 
bins of $\lvert\yz\rvert \in [0.0,\,1.0]$ and $\lvert\yz\rvert \in [1.0,\,2.1]$, and with an invariant-mass 
window of $80 < m_{\Pl\Pl} < 100~\GeV$ on the lepton-pair final state.
In the following, we perform a comparison to this CMS data for the measured $A_{0,..,4}$ coefficients.
For both rapidity selections, this comparison is performed for six bins within the range \ptz $\in [10,\,200]~\GeV$ 
as well as an overflow bin for $\ptz > 200~\GeV$.

\begin{figure}
\centering
\includegraphics[width=.49\linewidth]{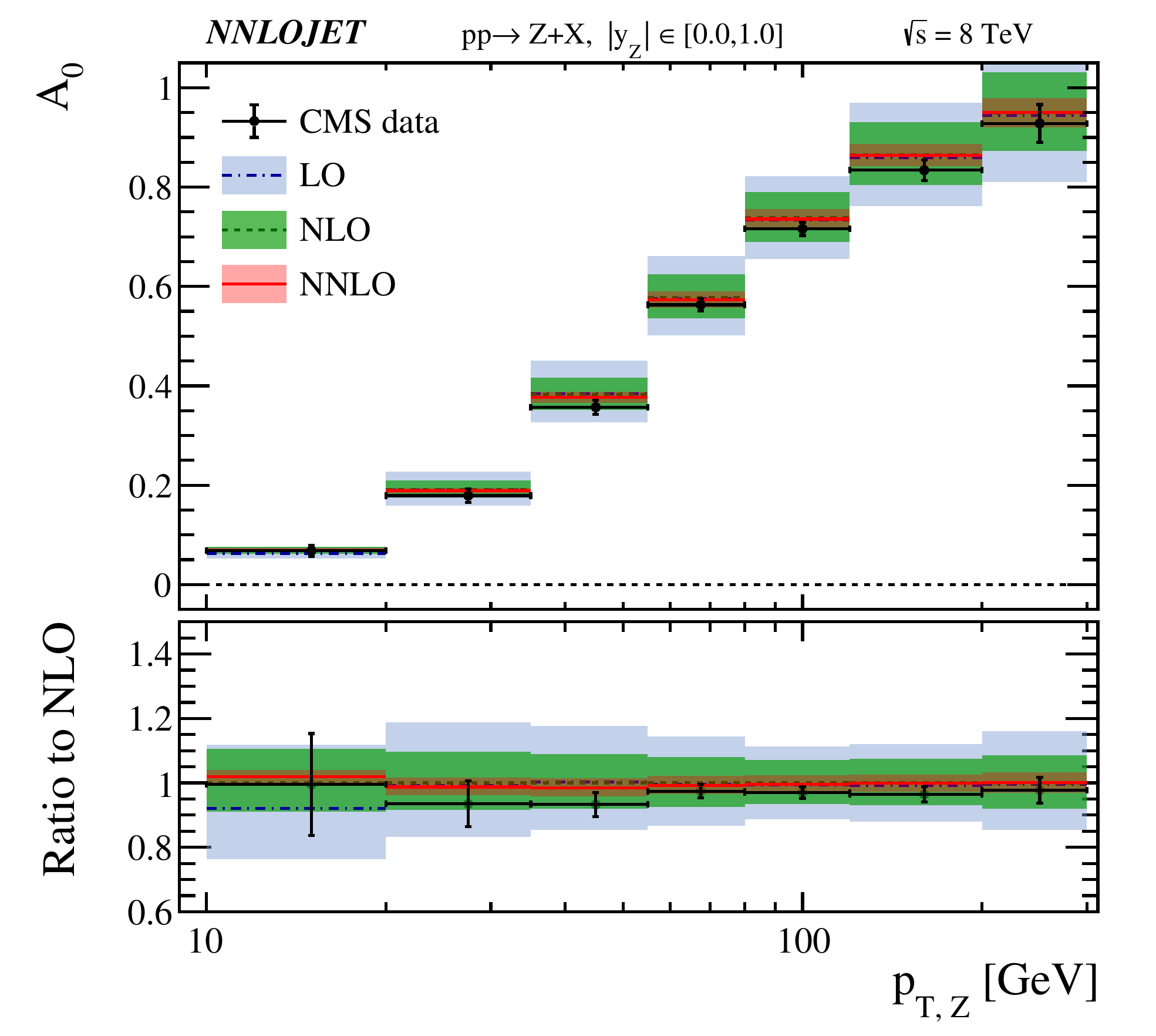} \hfill
\includegraphics[width=.49\linewidth]{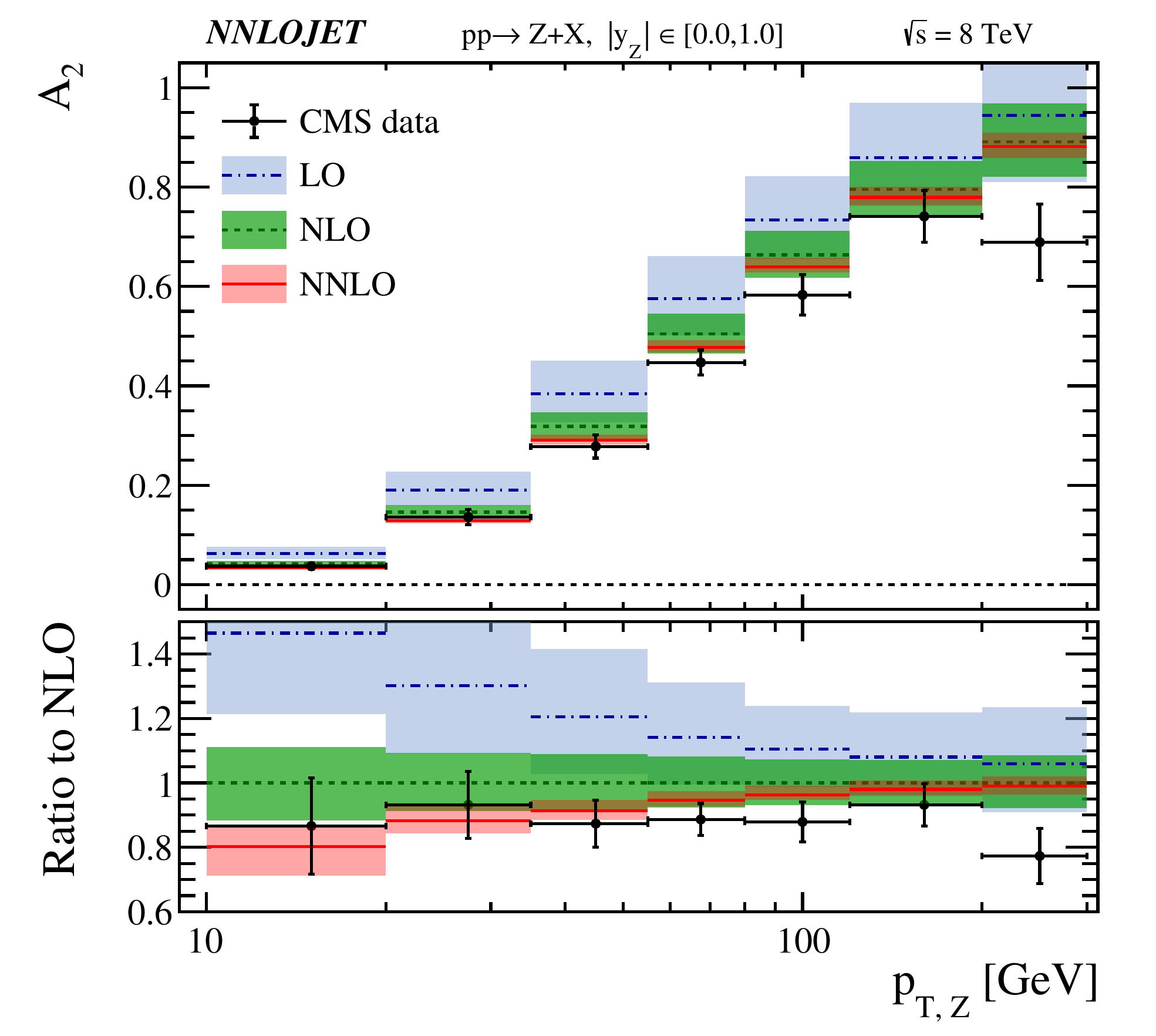} \\
\includegraphics[width=.49\linewidth]{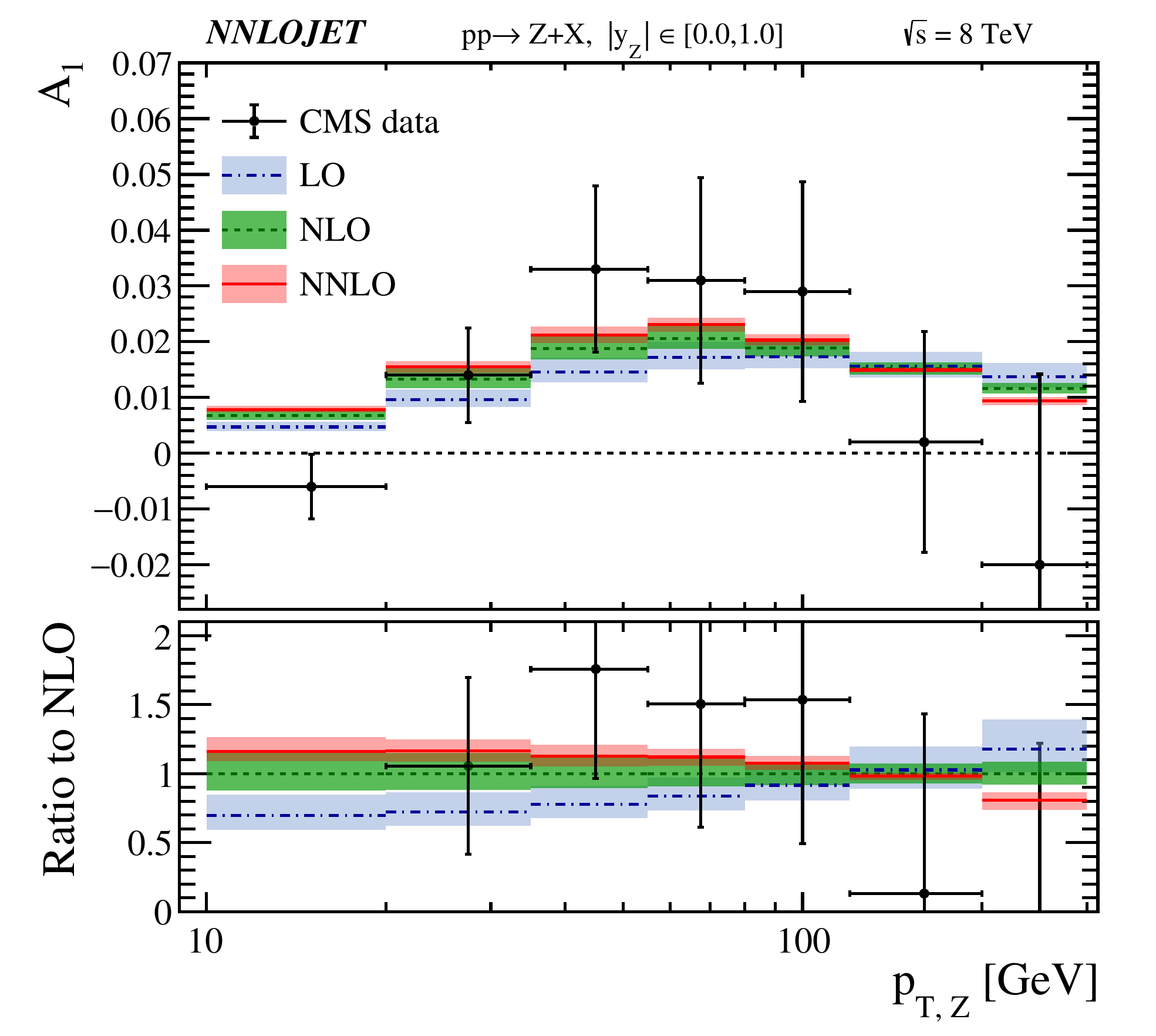} \hfill
\includegraphics[width=.49\linewidth]{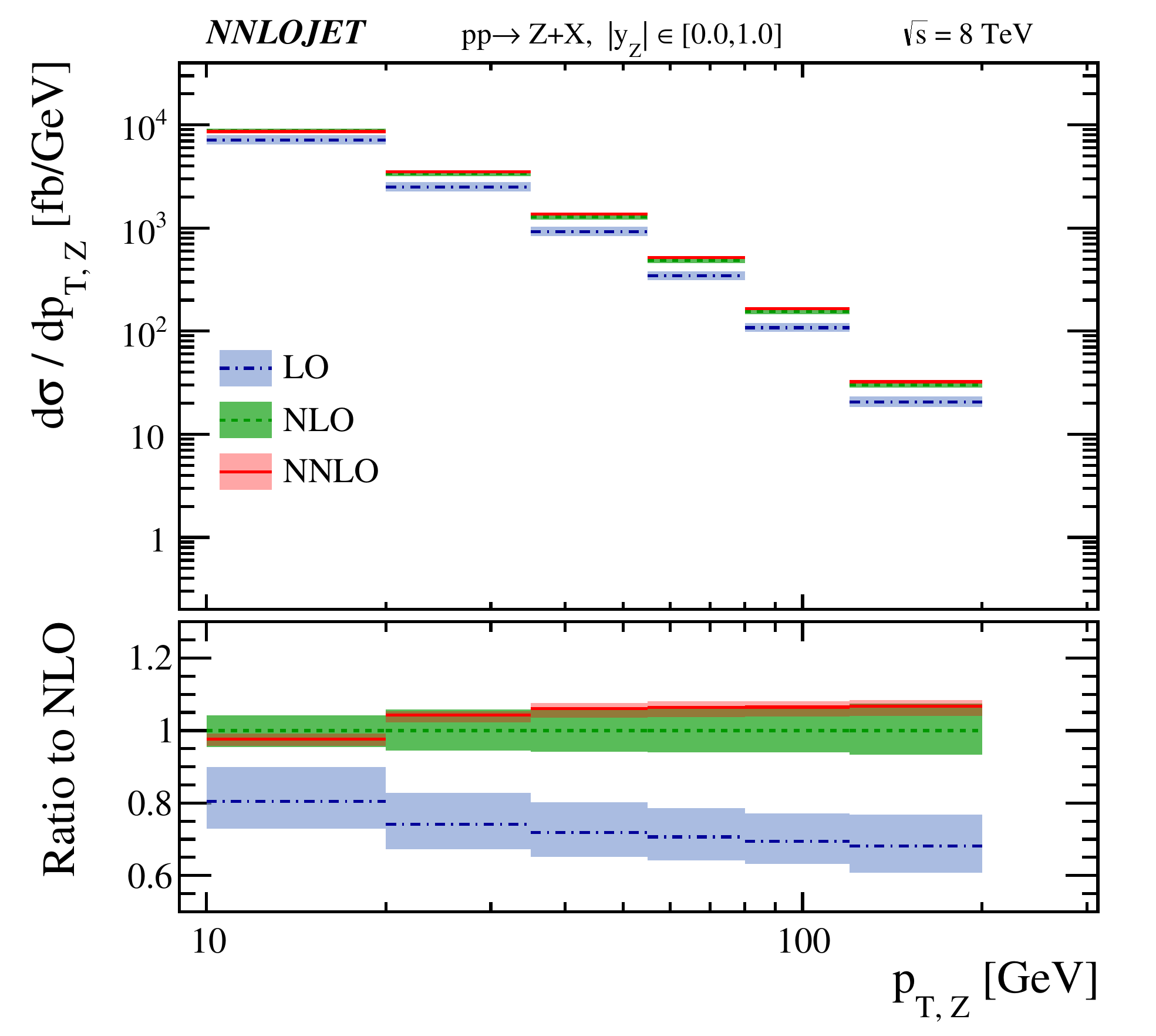}
\caption{
The \ptz distribution for the angular coefficients $A_0$ (upper left), $A_2$ (upper right),
$A_1$ (lower left), as well as the unpolarised cross section (lower right) in $\Pp\Pp$ collisions at $\sqrt{s} = 8~\TeV$
where a kinematic cut of $\lvert\yz\rvert \in [0.0,\,1.0]$ is required for all distributions.
The CMS data (black points) are compared to the LO (blue fill), NLO (green fill), and NNLO (red fill) 
theoretical predictions. In the lower panel, each distribution is shown normalised 
with respect to the central NLO prediction.
}
\label{fig:CMS_1}
\end{figure}
\begin{figure}
\centering
\includegraphics[width=.49\linewidth]{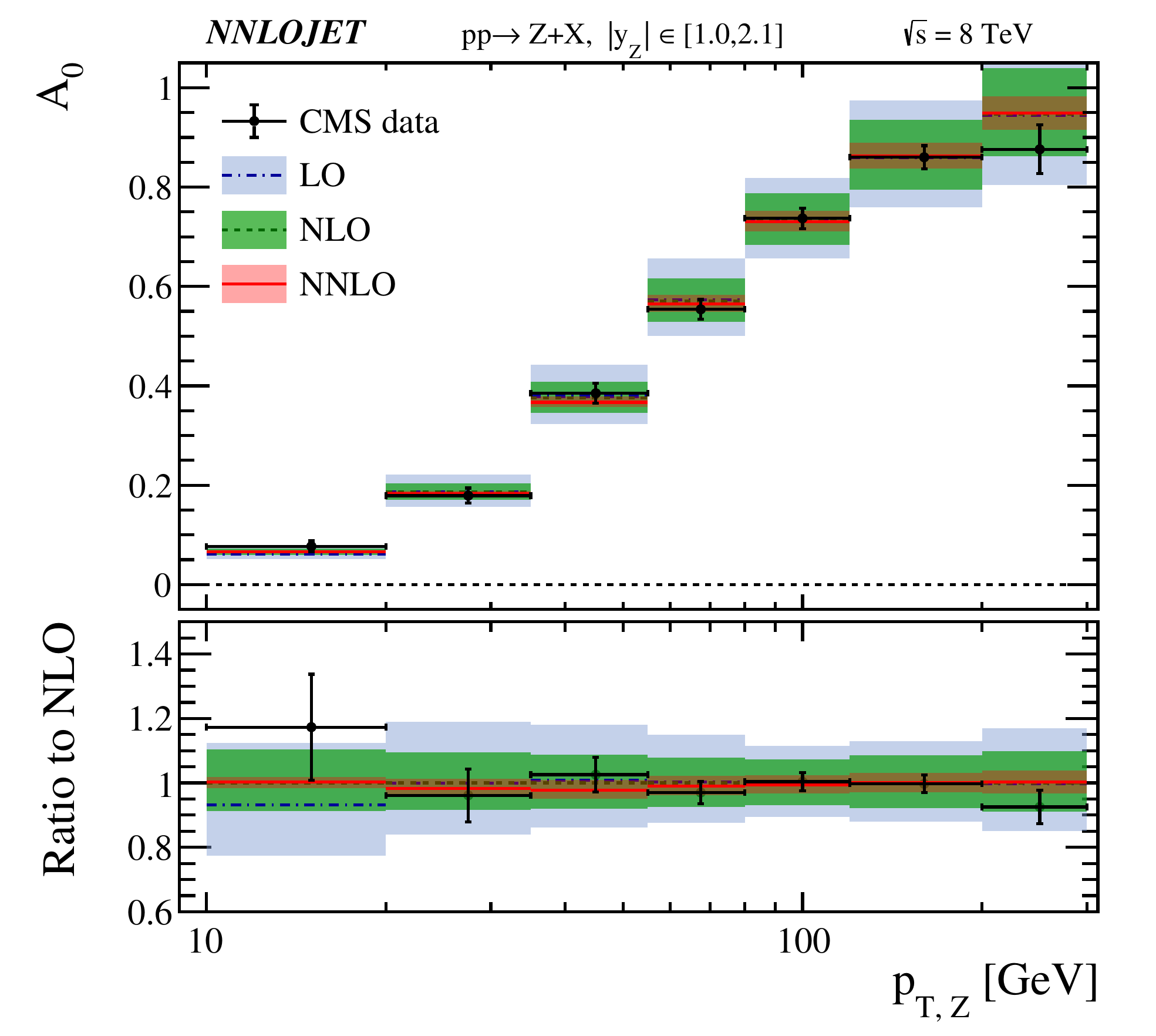} \hfill
\includegraphics[width=.49\linewidth]{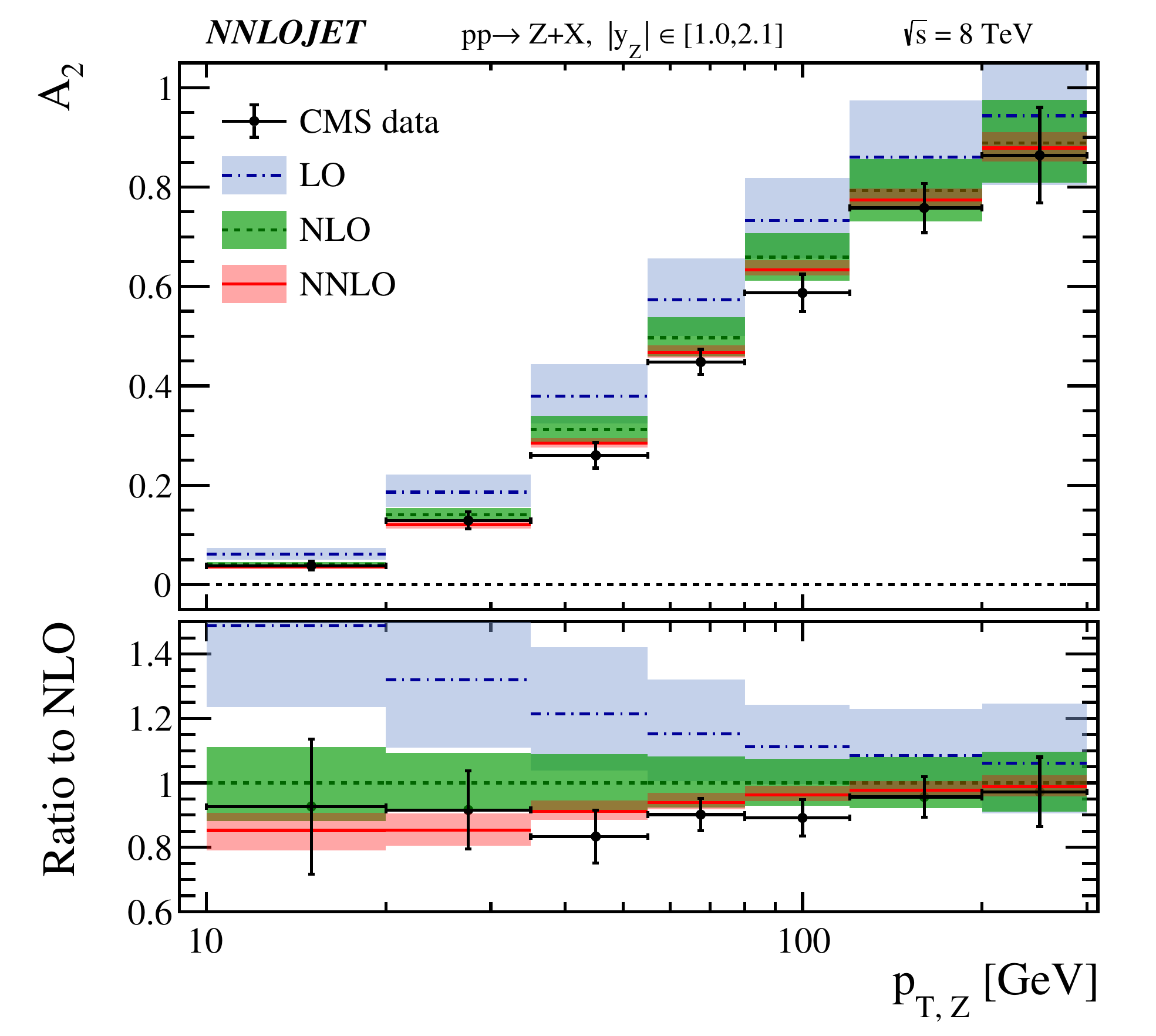} \\
\includegraphics[width=.49\linewidth]{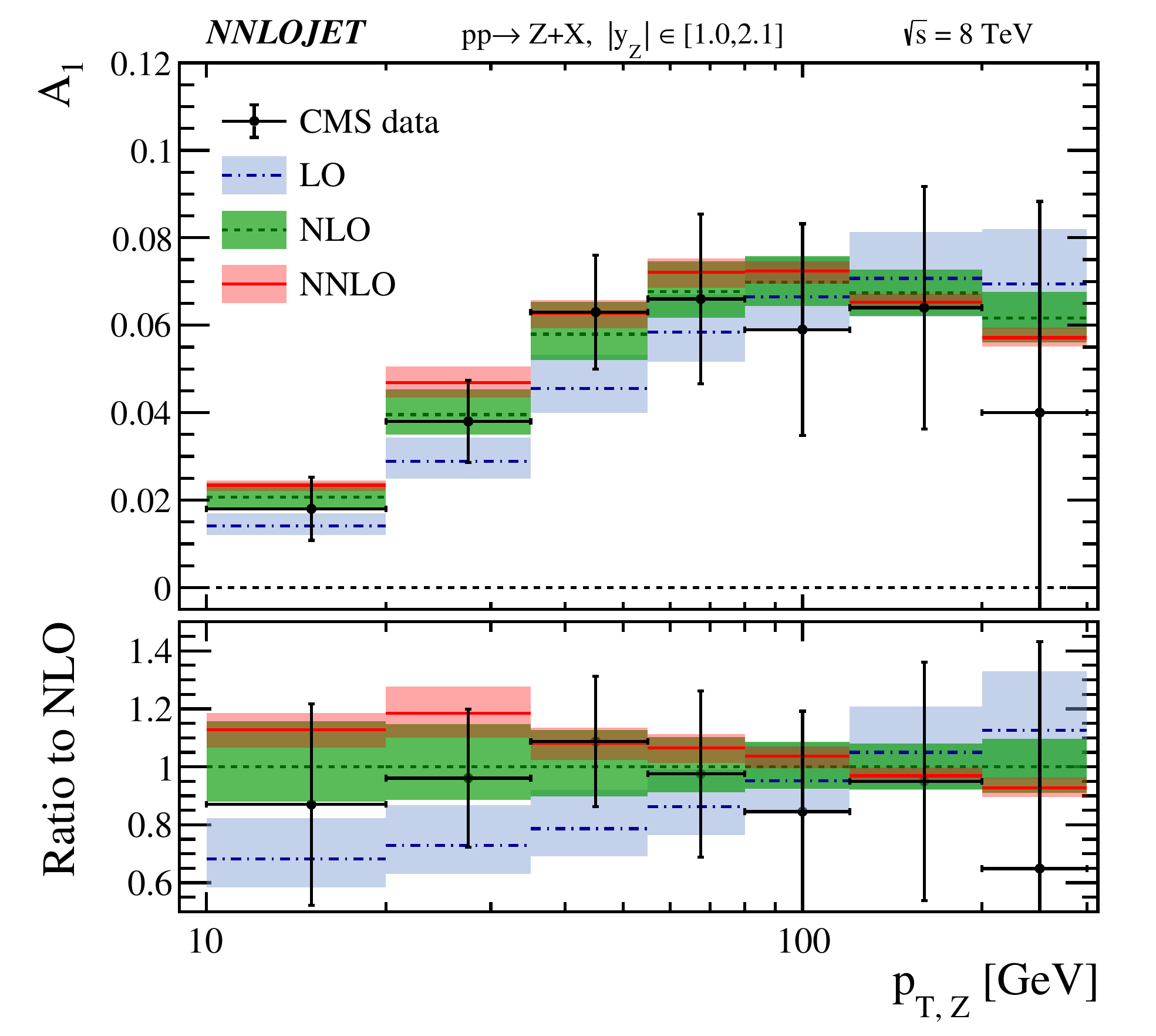} \hfill
\includegraphics[width=.49\linewidth]{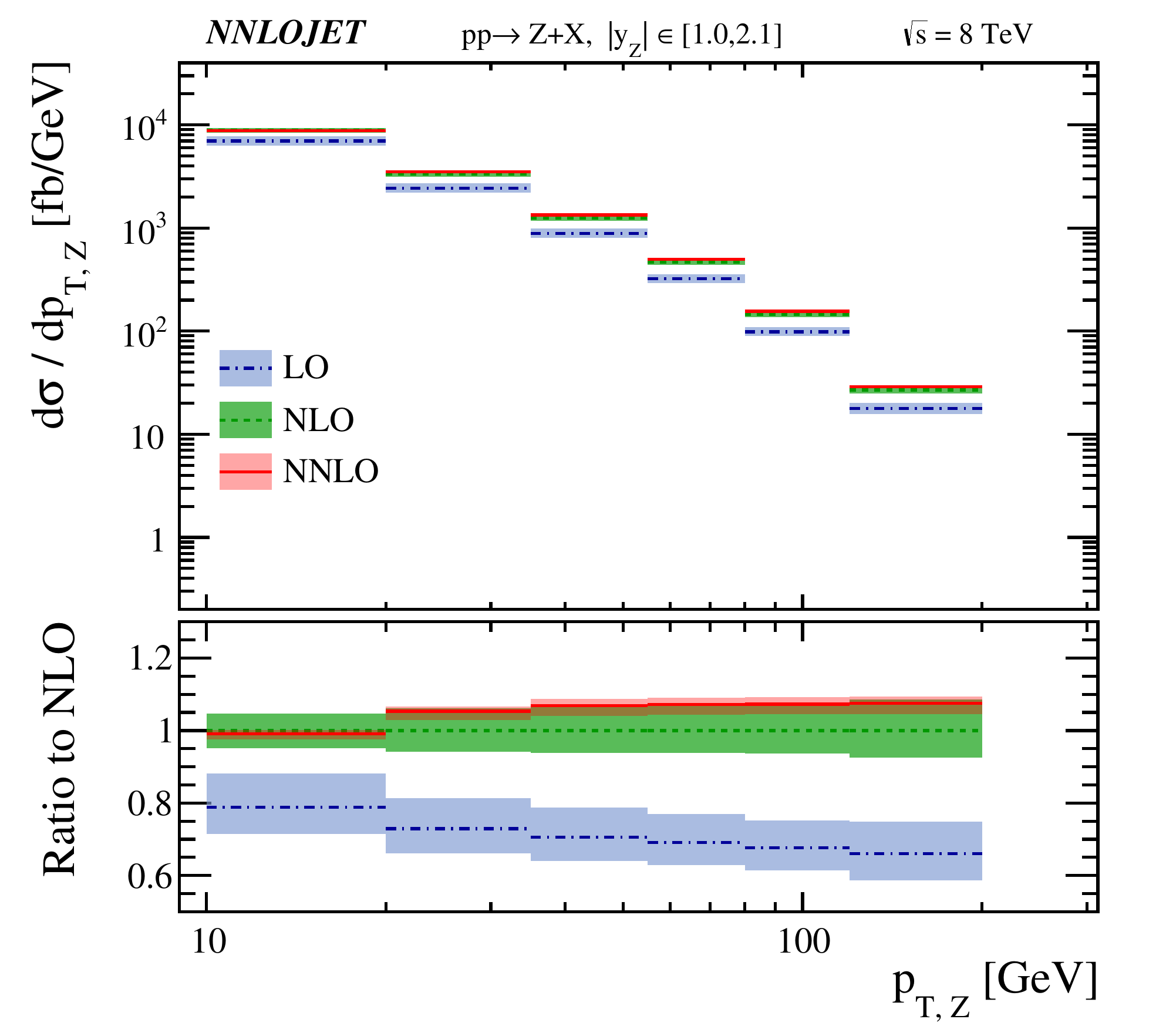}
\caption{
The \ptz distribution for the angular coefficients $A_0$ (upper left), $A_2$ (upper right),
$A_1$ (lower left), as well as the unpolarised cross section (lower right) in $\Pp\Pp$ collisions at $\sqrt{s} = 8~\TeV$
where a kinematic cut of $\lvert\yz\rvert \in [1.0,\,2.1]$ is required for all distributions.
The CMS data (black points) are compared to the LO (blue fill), NLO (green fill), and NNLO (red fill) 
theoretical predictions. In the lower panel, each distribution is shown normalised 
with respect to the central NLO prediction.
}
\label{fig:CMS_2}
\end{figure}

The distributions for $A_0$ (upper left), $A_1$ (lower left), $A_2$ (upper right), and the unpolarised cross section (lower right) are shown in Figs.~\ref{fig:CMS_1} and~\ref{fig:CMS_2}, where Fig.~\ref{fig:CMS_1} corresponds to the rapidity bin of $\lvert\yz\rvert \in [0.0,\,1.0]$, and Fig.~\ref{fig:CMS_2} to $\lvert\yz\rvert \in [1.0,\,2.1]$. The CMS data is represented by black points
and is compared to LO (blue), NLO (green), and NNLO (red) predictions. As before, 
the distributions are shown normalised to the central NLO prediction in the lower panel of each plot.
The NNLO corrections exhibit similar behaviour in both rapidity bins as was the case for
the rapidity-integrated distributions shown in Fig.~\ref{fig:ATLAS_1} for ATLAS. Namely, large negative
corrections (reaching $-15\%$) to $A_2$ are found within the range of \ptz $\in[20,\,100]~\GeV$, 
and positive (negative) corrections are observed in the $A_1$ distribution at low (large) \ptz.
The description of the data is visibly improved by the precise NNLO predictions, which
have relative scale uncertainties of order $5\%$.

\begin{figure}
\centering
\includegraphics[width=.49\linewidth]{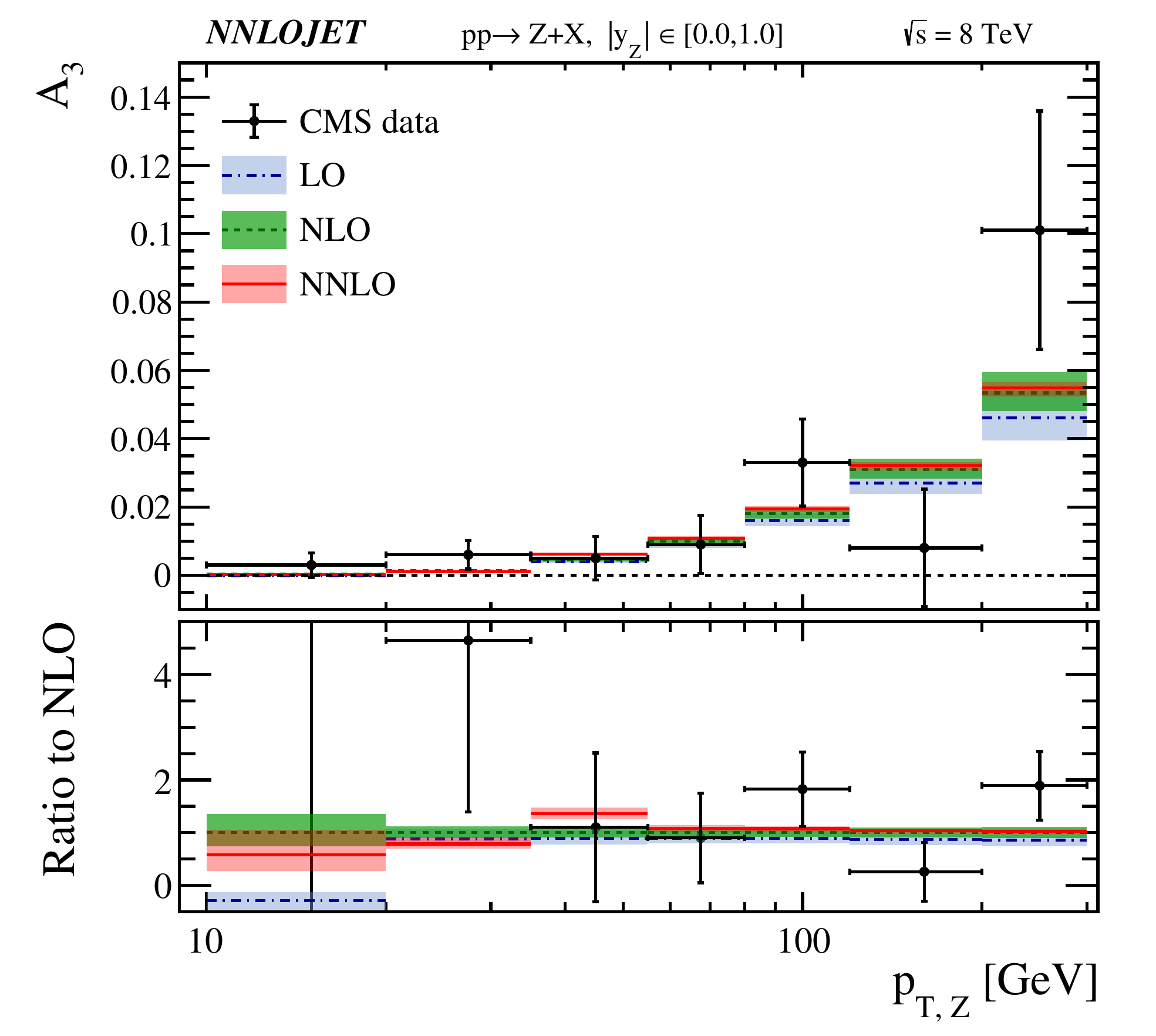} \hfill
\includegraphics[width=.49\linewidth]{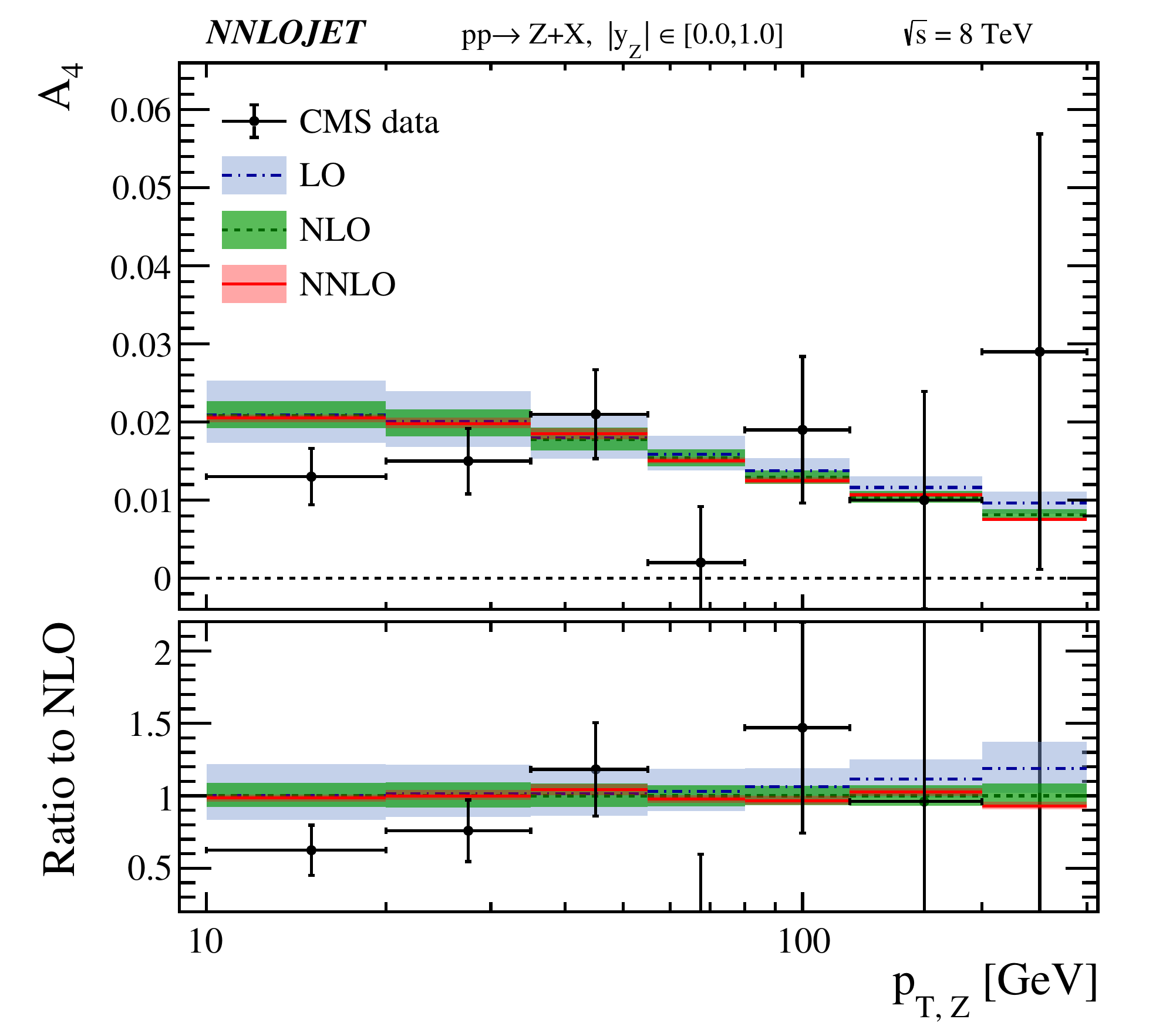} \\
\includegraphics[width=.49\linewidth]{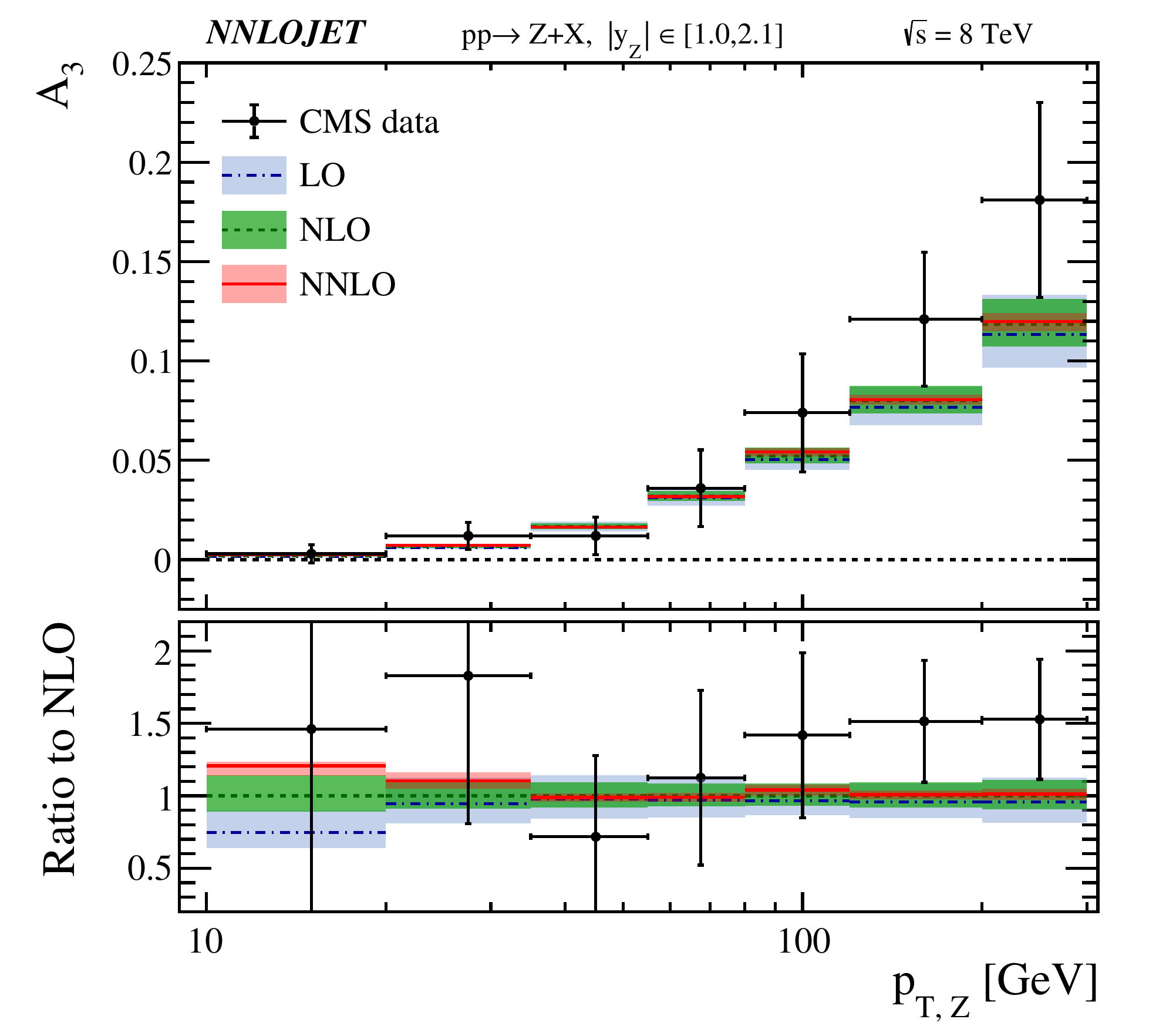} \hfill
\includegraphics[width=.49\linewidth]{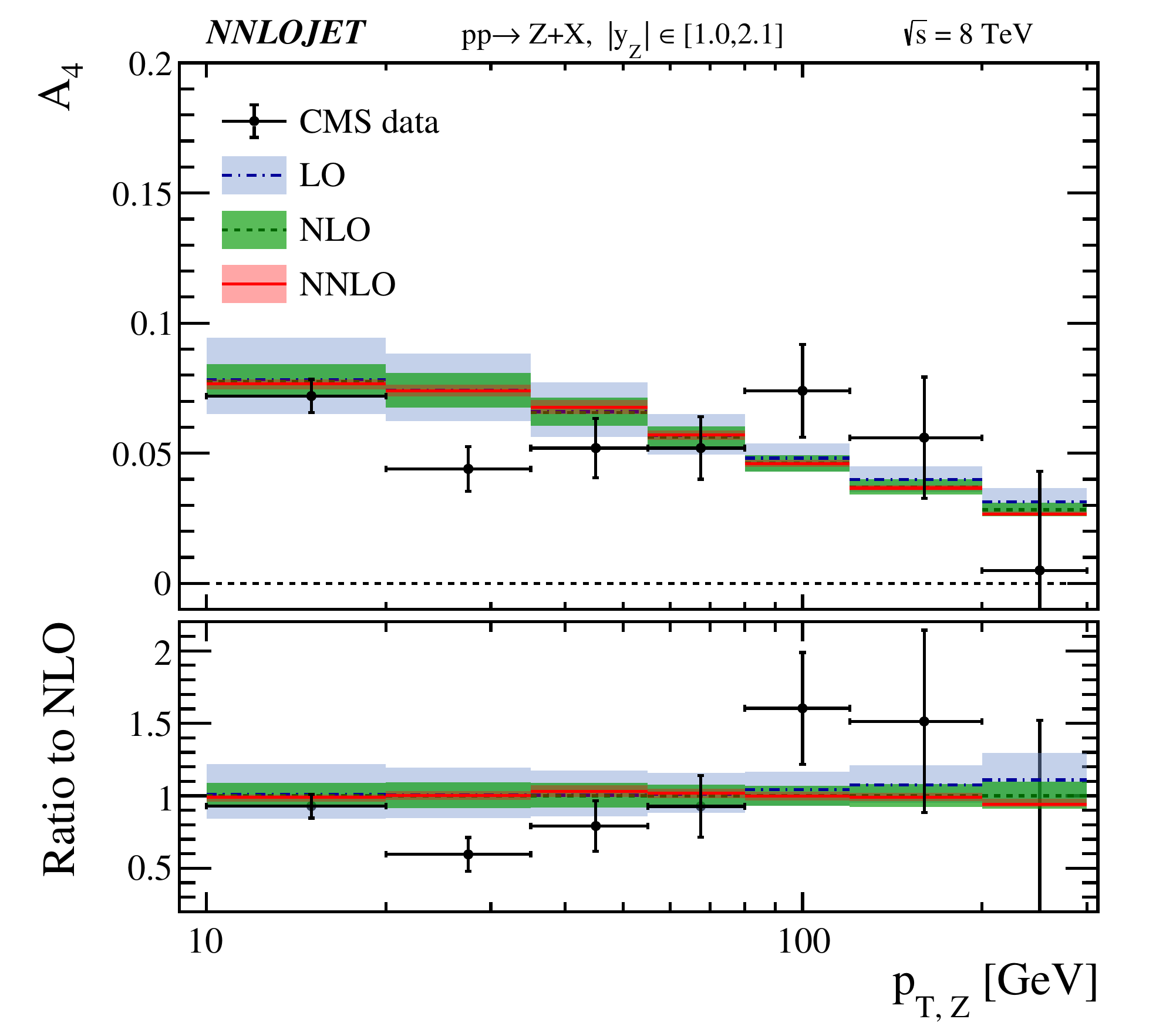}
\caption{
The \ptz distribution for the angular coefficients $A_3$ (left) and $A_4$ (right) 
in $\Pp\Pp$ collisions at $\sqrt{s} = 8~\TeV$, where kinematic cuts of $\lvert\yz\rvert \in [0.0,\,1.0]$ (left)
and $\lvert\yz\rvert \in [1.0,\,2.1]$ have been required for the shown distributions.
The CMS data (black points) are compared to the LO (blue fill), NLO (green fill), and NNLO (red fill) 
theoretical predictions. In the lower panel, each distribution is shown normalised 
with respect to the central NLO prediction. 
}
\label{fig:CMS_3}
\end{figure}

For the parity-violating angular coefficients $A_3$ and $A_4$, we see that the NNLO corrections do not alter the shapes of these distributions for the CMS kinematical setup (as was the case for the \yz-inclusive distributions for ATLAS shown in Fig. 5). Fig.~\ref{fig:CMS_3} shows that these distributions are well approximated by the central NLO predictions. The NNLO corrections to these distributions, as compared to the accuracy of the data, are phenomenologically unimportant.

\subsection{Predictions for LHCb}
The LHCb collaboration has not yet performed a measurement of the angular
coefficients in $\PZ$-boson production. Such an analysis would however be of interest to provide a
probe of the $\PZ$-boson production mechanisms which may be enhanced at forward rapidities, and is also an
important stepping-stone towards performing an extraction of $M_{\PW}$ within the LHCb acceptance~\cite{Bozzi:2015zja}. 
\begin{figure}
\centering
\includegraphics[width=.49\linewidth]{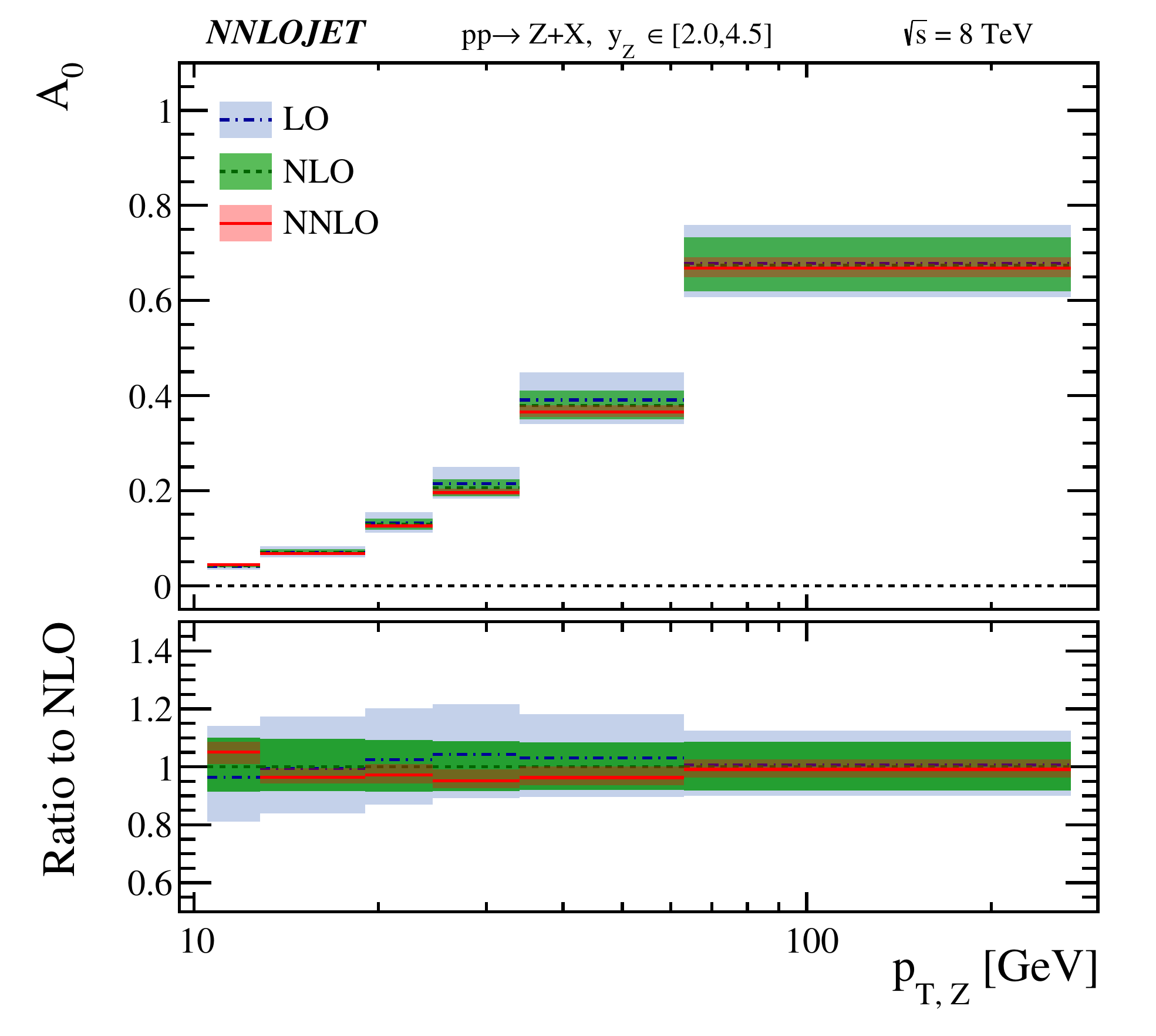} \hfill
\includegraphics[width=.49\linewidth]{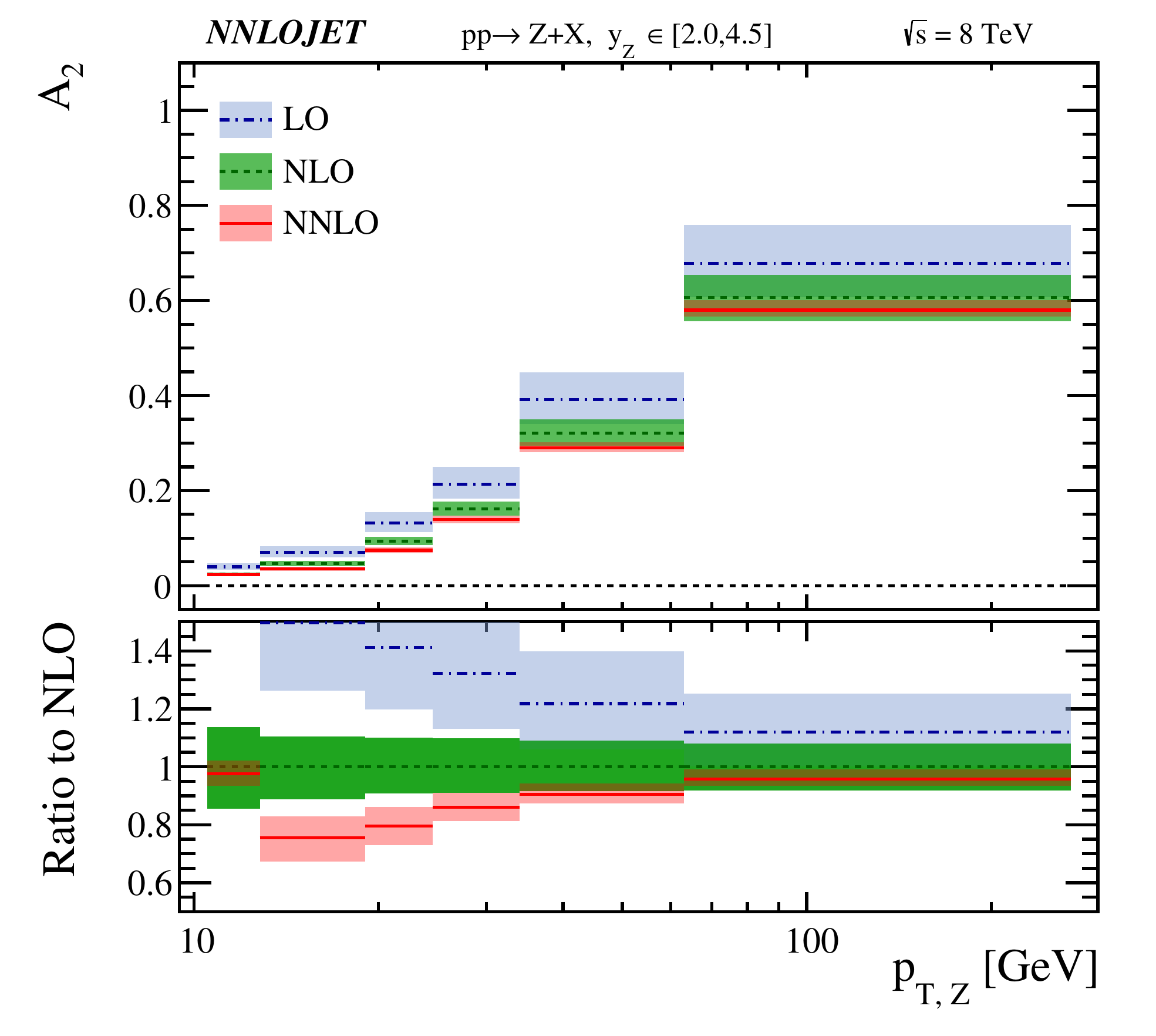} \\
\includegraphics[width=.49\linewidth]{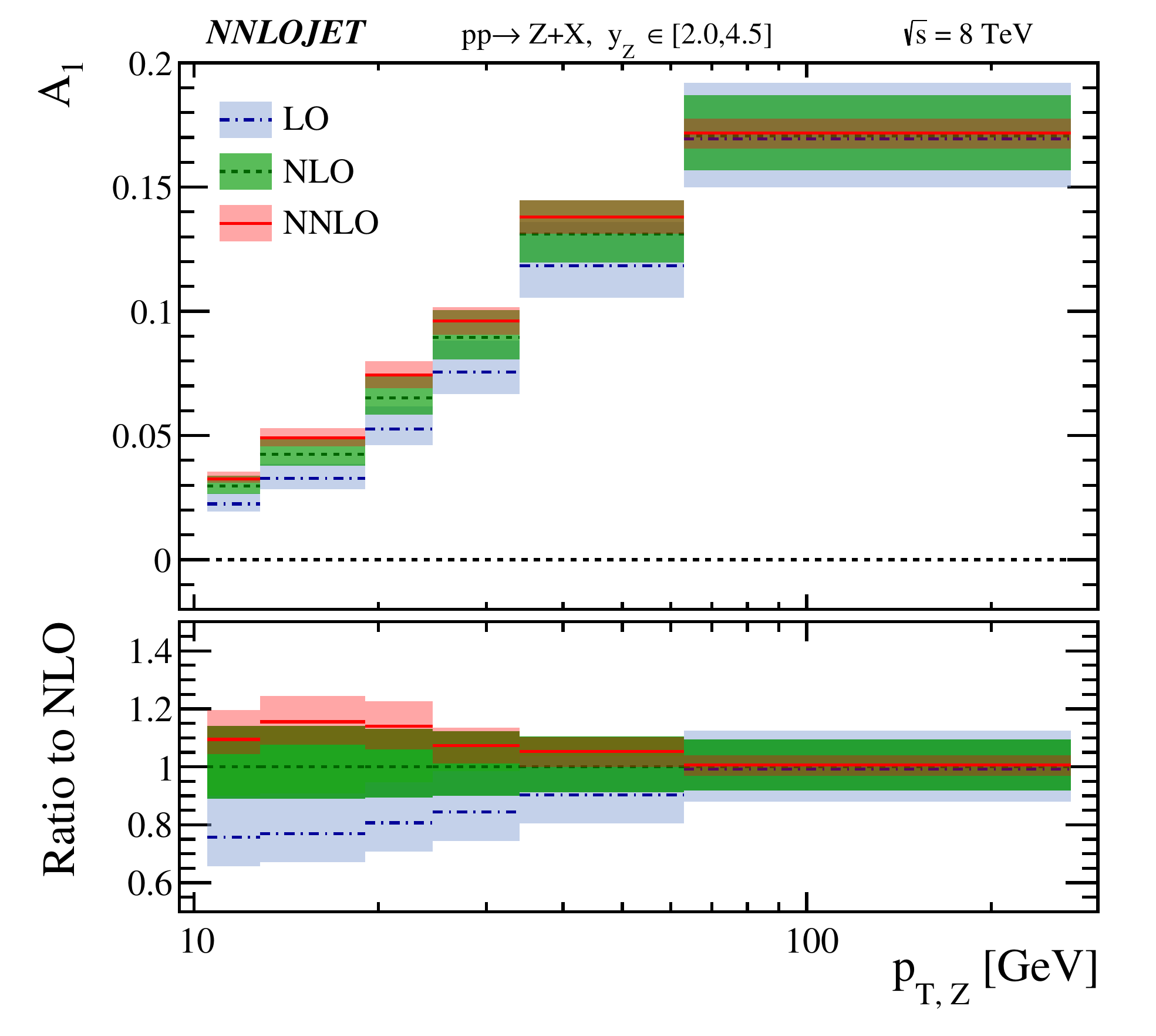} \hfill
\includegraphics[width=.49\linewidth]{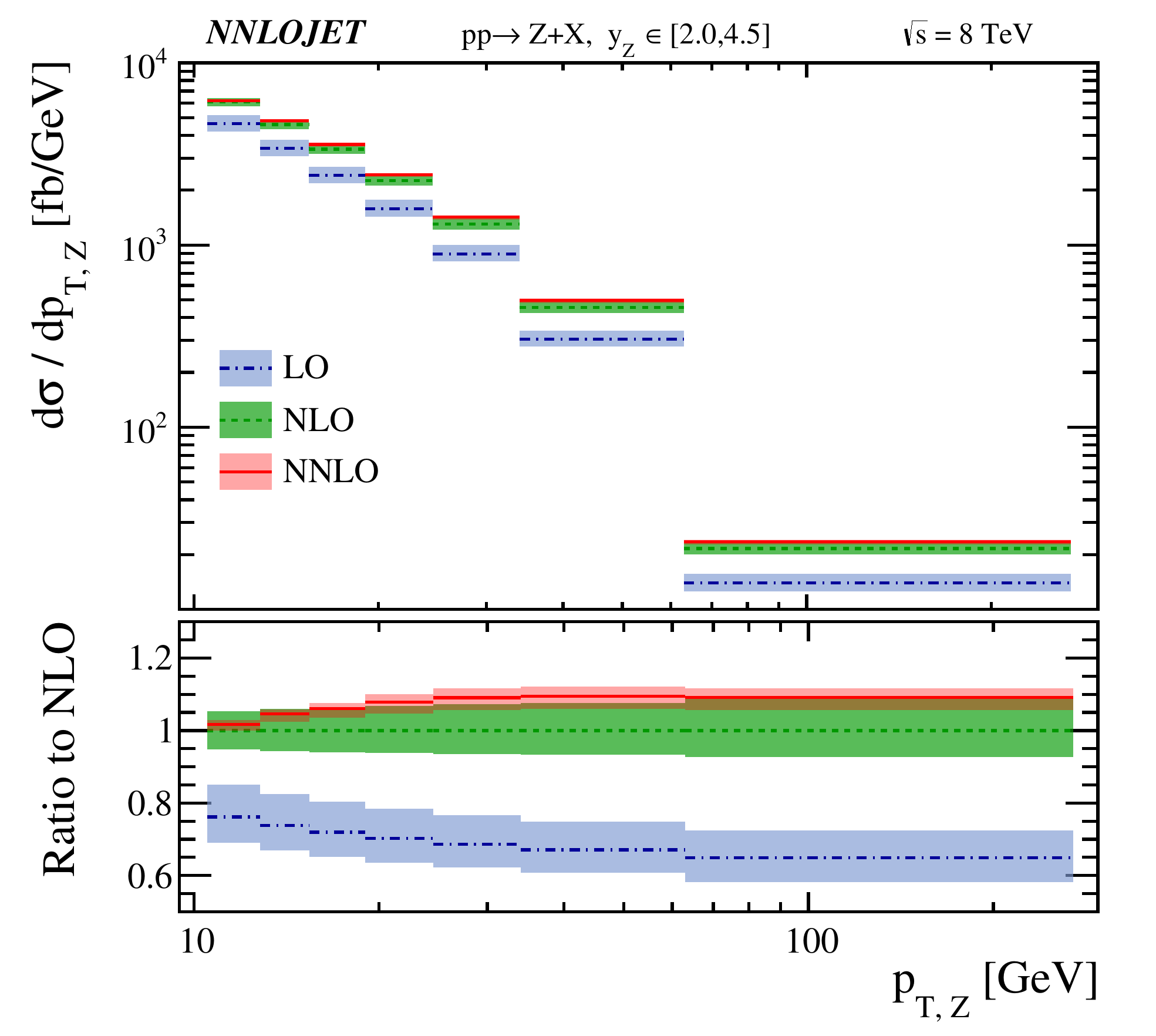}
\caption{\small 
The \ptz distribution for the angular coefficients $A_0$ (upper left), $A_2$ (upper right),
$A_1$ (lower left), as well as the unpolarised cross section (lower right) in $\Pp\Pp$ collisions at $\sqrt{s} = 8~\TeV$.
A kinematic cut of $\yz \in [2.0,\,4.5]$ (corresponding to the LHCb fiducial region) is required for all distributions.
Theoretical predictions are provided at LO (blue fill), NLO (green fill), and NNLO (red fill) 
accuracy. In the lower panel, each distribution is shown normalised 
with respect to the central NLO prediction.
}
\label{fig:LHCb_1}
\end{figure}
We therefore provide predictions for the $\ptz$ distributions for the LHCb fiducial region of 
\yz $\in [2.0,\,4.5]$, placing an invariant mass selection of $80 < m_{\Pl\Pl} < 100$~GeV on the lepton-pair final state.
No other cuts are placed on the lepton-pair final state as it is assumed that this will be corrected for in 
the experimental analysis.
The predictions for $A_0$ (upper left), $A_1$ (lower left), $A_2$ (upper right), as well as
the unpolarised cross section (lower right) are provided in Fig.~\ref{fig:LHCb_1}. Each \ptz distribution 
is provided in the region $\ptz \in [10.5,\,270]$~GeV, guided by the choice of binning taken 
in the recent LHCb measurement of $\PZ$-boson production at 13~TeV~\cite{Aaij:2016mgv}.

The predicted shapes of the distributions within the LHCb acceptance are similar
to what is observed at more central rapidities. 
In addition, the NNLO corrections 
for each of the angular coefficients are also observed to be of similar size to those
at more central rapidities. Although not shown here, the NNLO corrections to $A_3$ and $A_4$
were also computed for this kinematic setup and found to be negligibly small.
We can therefore conclude that the NNLO corrections to the \ptz spectrum 
for $A_0$, $A_1$, and $A_2$ should be included when performing a comparison to data, while
the central NLO prediction for the corresponding $A_3$ and $A_4$ distributions are likely sufficient.
If an experimental determination of the $A_3$ and $A_4$ distributions is achievable at LHCb
with smaller relative uncertainties as compared to ATLAS and CMS, it would be important to include 
the effects of electroweak corrections and to also assess the impact of PDF uncertainties 
on these distributions.

\subsection{Assessing the violation of the Lam--Tung relation} 
\label{LamTung}
As highlighted in Section~\ref{sec2}, the Lam--Tung relation is expected to be violated, i.e.\ $(A_0-A_2)\ne0$,
starting at order $\cO(\alphas^2)$ in the framework of pQCD.
Measurements of this violation therefore provide an important test of the
$\PZ$-boson production dynamics. As discussed in Section~\ref{sec2}, 
the extent of the breaking of the Lam--Tung relation can be assessed by measuring 
the \ptz distribution for the difference of the angular coefficients $A_0$ and $A_2$, or equivalently through
the normalised observable $\Delta^{\rm LT}$. The latter has the benefit of better exposing 
the violation of the Lam--Tung relation in the lower \ptz range, where the angular 
coefficients $A_0$ and $A_2$ are individually relatively small.
In the following, we discuss the corrections to $(A_0-A_2)$ and quantify the consistency of the 
data and the predictions by performing a $\chi^2$ test. We then present results for $\Delta^{\rm LT}$ 
and perform a comparison to data, where the data points for the latter are obtained by 
re-expressing $\Delta^{\rm LT}$ in terms of the measured angular coefficients.

Before comparing to data, it is important to comment on the expected accuracy of our theoretical predictions for 
these two observables. As for the individual angular coefficients, our theoretical predictions for $(A_0-A_2)$ 
are obtained from the computation of the production process for $\PZ+\jet$ at $\cO(\alphas^3)$. 
While the $\cO(\alphas^3)$ contributions comprise genuine NNLO corrections to the individual $A_i$ coefficients 
as demonstrated throughout this section, the prediction degrades to an NLO-accurate description for the 
difference $(A_0-A_2)$ and $\Delta^{\rm LT}$.\footnote{In the sense that the first non-trivial prediction for these observables begins at $\cO(\alphas^2)$.}
For consistency with the rest of the paper, we will continue to refer to the corrections of order $\cO(\alphas^3)$ as 
``NNLO corrections'' and similarly label the figures in this section as NNLO predictions.

\begin{figure}
\centering
\includegraphics[width=.49\linewidth]{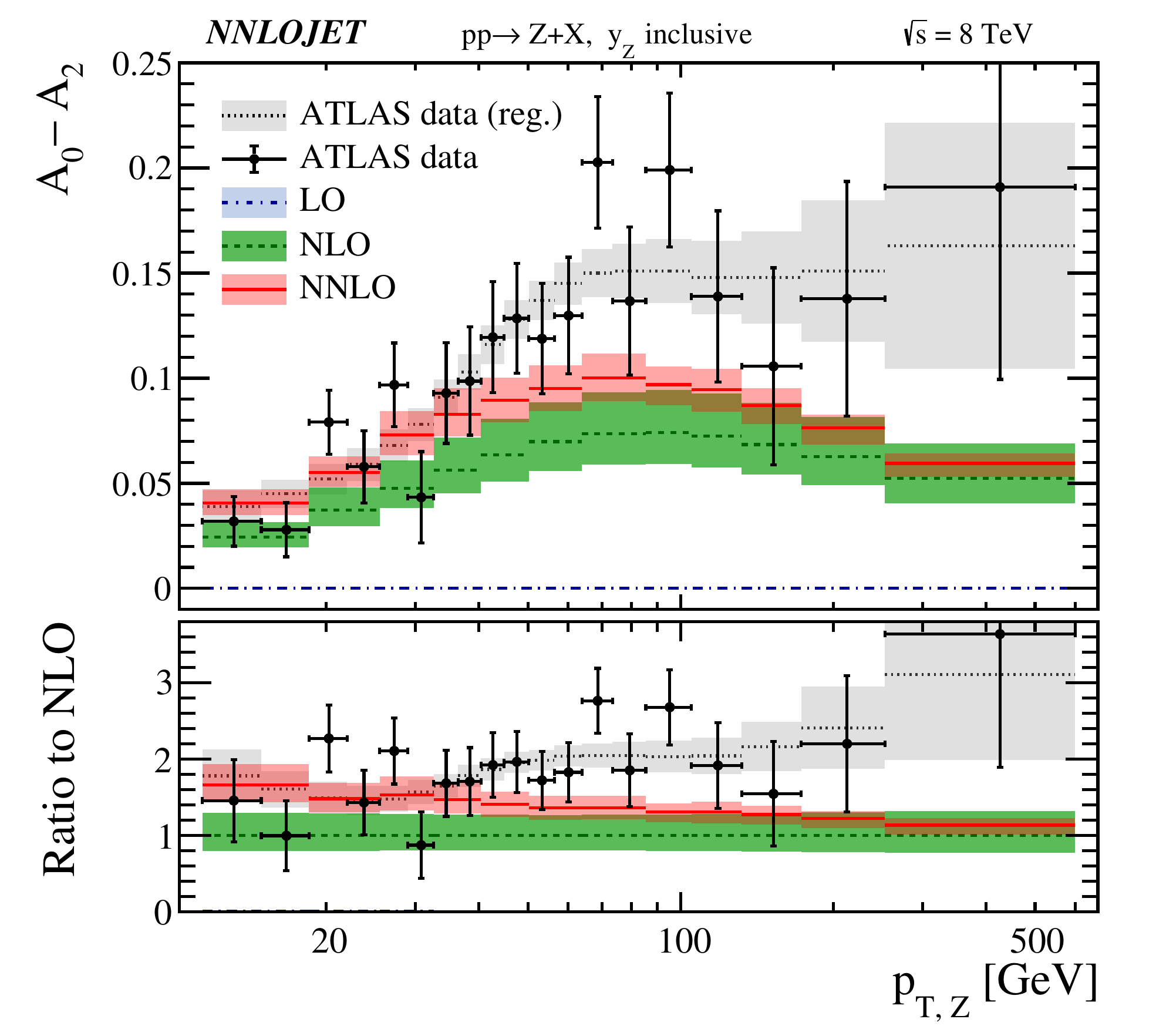}
\caption{\small 
The \ptz distribution for the difference of angular coefficients $(A_0-A_2)$ in $\Pp\Pp$ collisions at $\sqrt{s} = 8~\TeV$. 
The ATLAS data (black points) are compared to the LO (blue fill), NLO (green fill), and NNLO (red fill) 
theoretical predictions. In addition, the regularised ATLAS data is also included (grey fill). In the lower panel, 
each distribution is shown normalised with respect to the central NLO prediction.
}
\label{fig:ATLAS_A0mA2}
\bigskip
\includegraphics[width=.49\linewidth]{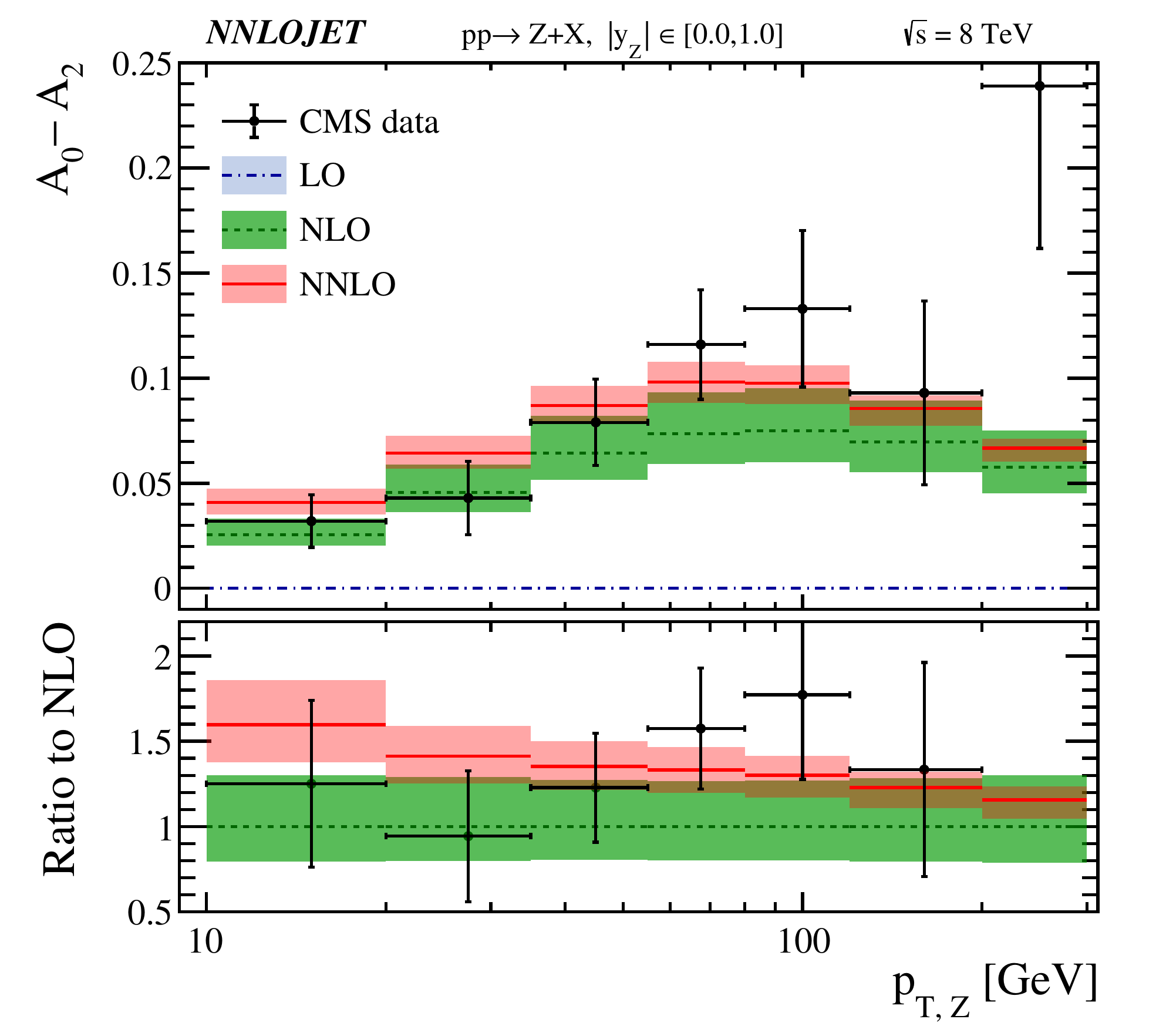} \hfill
\includegraphics[width=.49\linewidth]{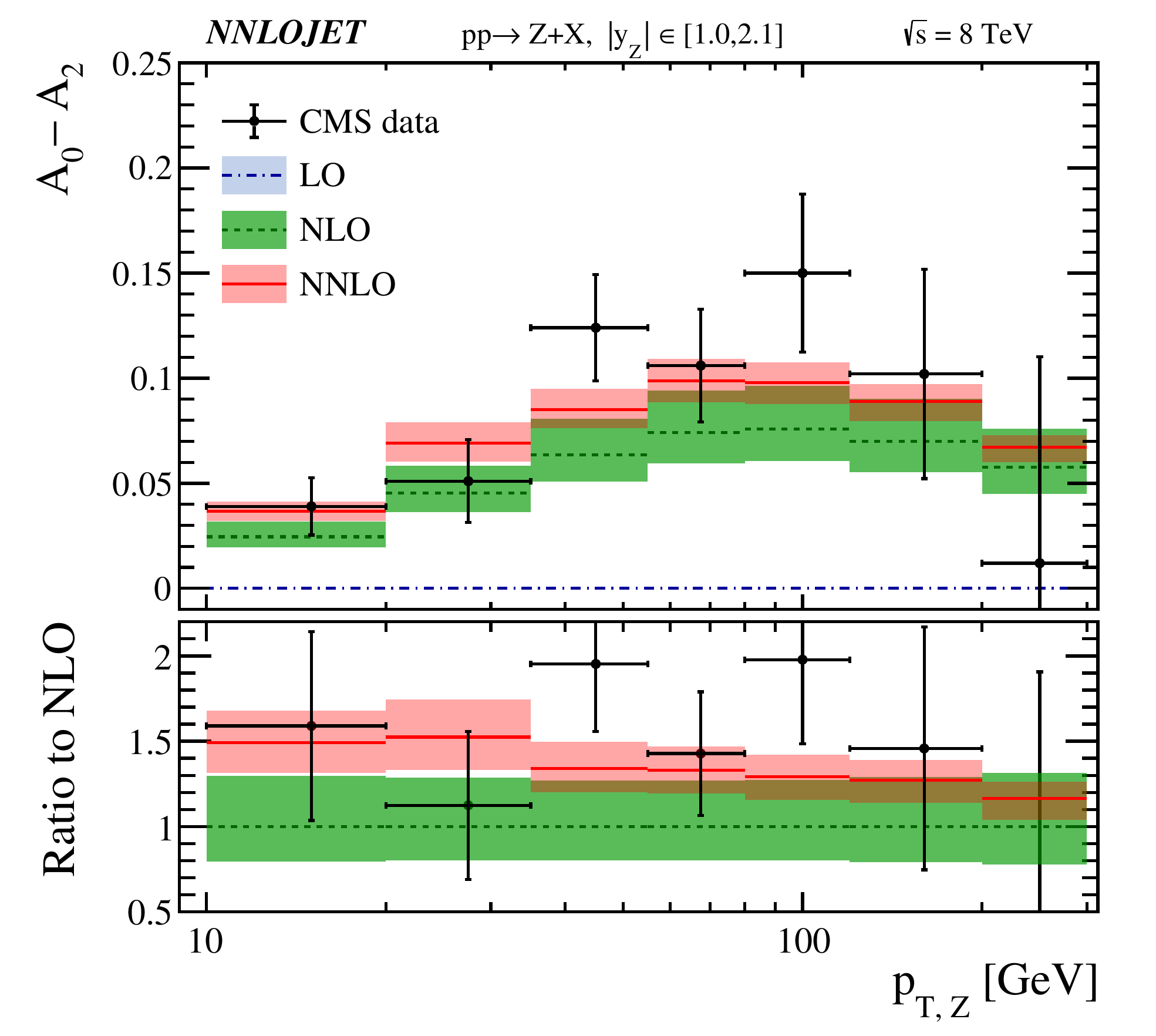}
\caption{\small
The \ptz distribution for the difference of angular coefficients $(A_0-A_2)$ in $\Pp\Pp$ collisions at $\sqrt{s} = 8~\TeV$,
with the kinematic cuts of $\lvert\yz\rvert \in [0.0,\,1.0]$ (left) and $\lvert\yz\rvert \in [1.0,\,2.1]$ (right). 
The CMS data (black points) are compared to the LO (blue fill), NLO (green fill), and NNLO (red fill) 
theoretical predictions. In the lower panel, each distribution is shown normalised with respect to the central NLO prediction.}
\label{fig:CMS_A0mA2}
\end{figure}

Figure~\ref{fig:ATLAS_A0mA2} shows the \ptz distribution for $(A_0-A_2)$,
where the ATLAS data is represented by black points, and is compared to LO (blue), 
NLO (green), and NNLO (red) theoretical predictions.
The NNLO corrections are observed to be large and positive, amounting to $+40\%$ at moderate $\ptz$ values,
and provide an improved description of the ATLAS data. It is worth noting that while a reduction of the absolute 
scale uncertainties is already observed at NNLO with respect to NLO, the relative uncertainty is in fact reduced 
by almost a factor of two across the shown $\ptz$ range.
This is a reflection of the fact that the computation of the $\PZ+\jet$-production process at
$\cO(\alphas^3)$ at finite $\ptz$ used to predict the difference $(A_0-A_2)$ is only NLO accurate  
and therefore yields corrections and remaining scale uncertainties which are typical for NLO
effects.
For most of the \ptz range, and for both experimental setups, it is found that the NLO and NNLO predictions 
for $(A_0-A_2)$ are consistent within uncertainties.
In Fig.~\ref{fig:ATLAS_A0mA2}, we have also chosen to include the regularised ATLAS data (indicated
by the grey fill). As discussed towards the start of this section, large bin-to-bin correlations are introduced
in the regularisation procedure of the ATLAS data (see for example Fig.~24 of Ref.~\cite{Aad:2016izn}).
We believe that this demonstrates how a visual comparison of the theory prediction with respect to the
regularised data (in this case at least) can lead one to overestimate the disagreement between theory and data.

As an alternative to a visual comparison, the quality of the theoretical description of the data can be
quantified by performing a $\chi^2$ test according to
\begin{align}
\chi^2 = \sum_{i,j}^{\rm N_{data}} ( O_{\rm exp}^i - O_{\rm th.}^i ) \sigma_{ij}^{-1} ( O_{\rm exp}^j - O_{\rm th.}^j ) ,
\end{align}
where $O_{\rm exp}^i$ and $O_{\rm th.}^i$ are respectively the central value of the experimental and theoretical predictions
for data point $i$, and $\sigma_{ij}^{-1}$ is the inverse covariance matrix.
In this case, we consider the unregularised ATLAS data for the angular coefficients $A_0$ and $A_2$ (and their correlations, as provided through the covariance matrix) within the range of $\ptz \in [11.4,\,600]~\GeV$, corresponding to a total of 38 data points. To perform this comparison, the central theory predictions for these angular coefficients are evaluated with the same binning choice as the data.
The results are
\begin{align} 
  \text{NLO  (ATLAS):}&\qquad \chi^2/{\rm N_{data}} = 185.8/38 = 4.89 \,, \nonumber \\
  \text{NNLO (ATLAS):}&\qquad \chi^2/{\rm N_{data}} = 68.3/38  = 1.80 \,. \nonumber
\end{align}
This test indeed demonstrates that the NLO predictions give a poor description of the
data in the considered $\ptz$ range, a point that was also highlighted in the experimental analysis~\cite{Aad:2016izn}. 
This tension is largely reduced with the inclusion of the NNLO corrections, and from closer inspection 
of Fig.~\ref{fig:ATLAS_1}, can be mainly attributed to the large negative corrections to the $A_2$ distribution.

The corresponding \ptz distributions for the CMS measurement are shown in Fig.~\ref{fig:CMS_A0mA2}
for the rapidity bins $\lvert\yz\rvert \in [0.0,\,1.0]$ (left) and $\lvert\yz\rvert \in [1.0,\,2.1]$ (right). 
The NNLO corrections to $(A_0-A_2)$ exhibit a similar behaviour for the CMS kinematic selections,
and again improve the description of data. This agreement can also be quantified by performing 
a $\chi^2$ test, where in this case the test is performed directly on the $(A_0-A_2)$ distribution 
as no covariance matrix for these $A_i$ coefficients is publicly available. In total 14 data points are considered, corresponding
to seven \ptz bins for each rapidity selection. The results, assuming uncorrelated bins, are
\begin{align} 
  \text{NLO  (CMS):}&\qquad \chi^2/{\rm N_{data}} = 24.5/14 = 1.75\,, \nonumber \\
  \text{NNLO (CMS):}&\qquad \chi^2/{\rm N_{data}} = 14.2/14 = 1.01 \,. \nonumber
\end{align}
Similar to the findings for the ATLAS data, the description of the CMS data is substantially improved at NNLO.

\begin{figure}[ht!]
\centering
\includegraphics[width=.49\linewidth]{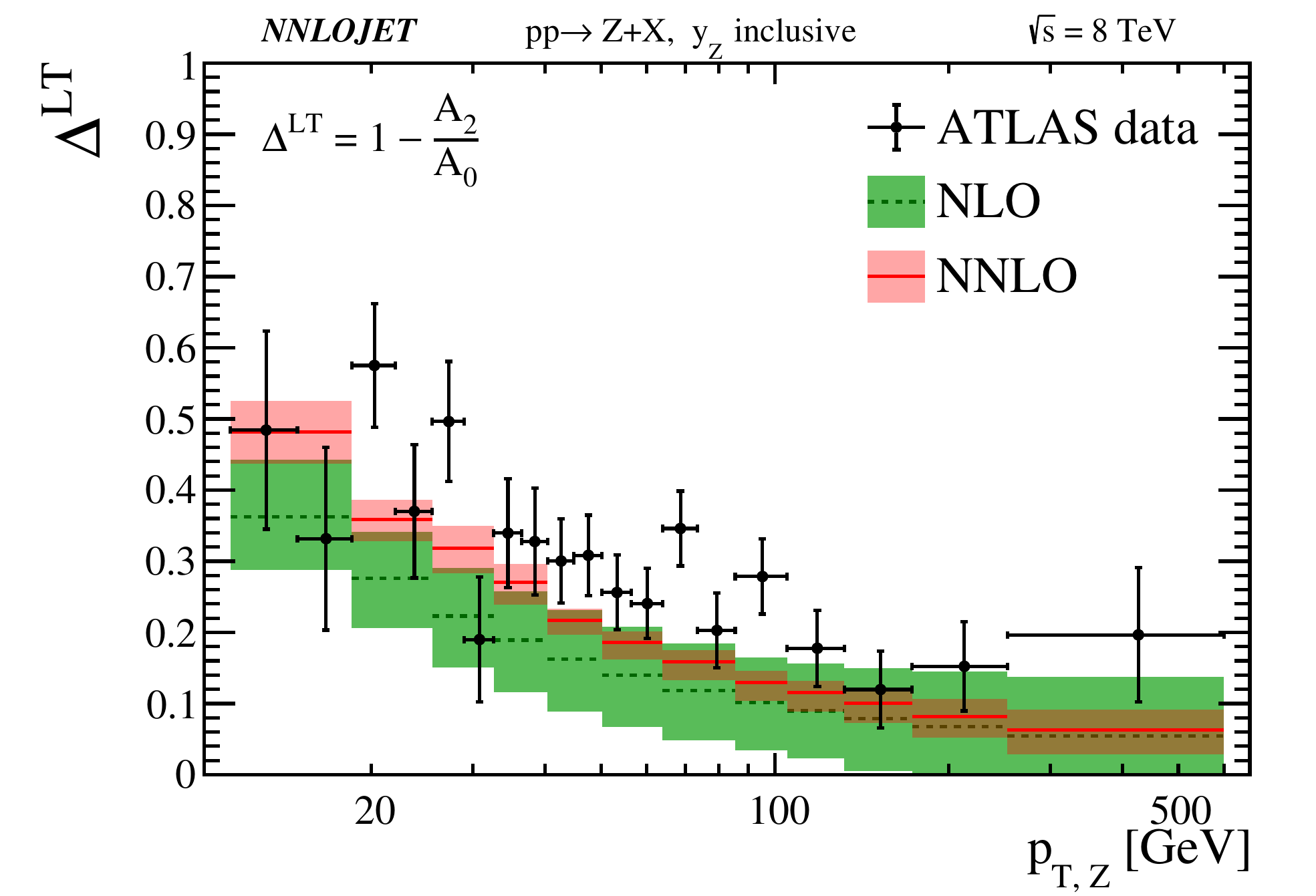}
\caption{\small The extent of the Lam--Tung violation as expressed through $\Delta^{\rm LT}$
for the ATLAS data in $\Pp\Pp$ collisions at $\sqrt{s} = 8~\TeV$. The data is compared to the corresponding 
NLO (green fill) and NNLO (red fill) predictions.
}
\label{fig:LamTungATLAS}
\bigskip
\includegraphics[width=.49\linewidth]{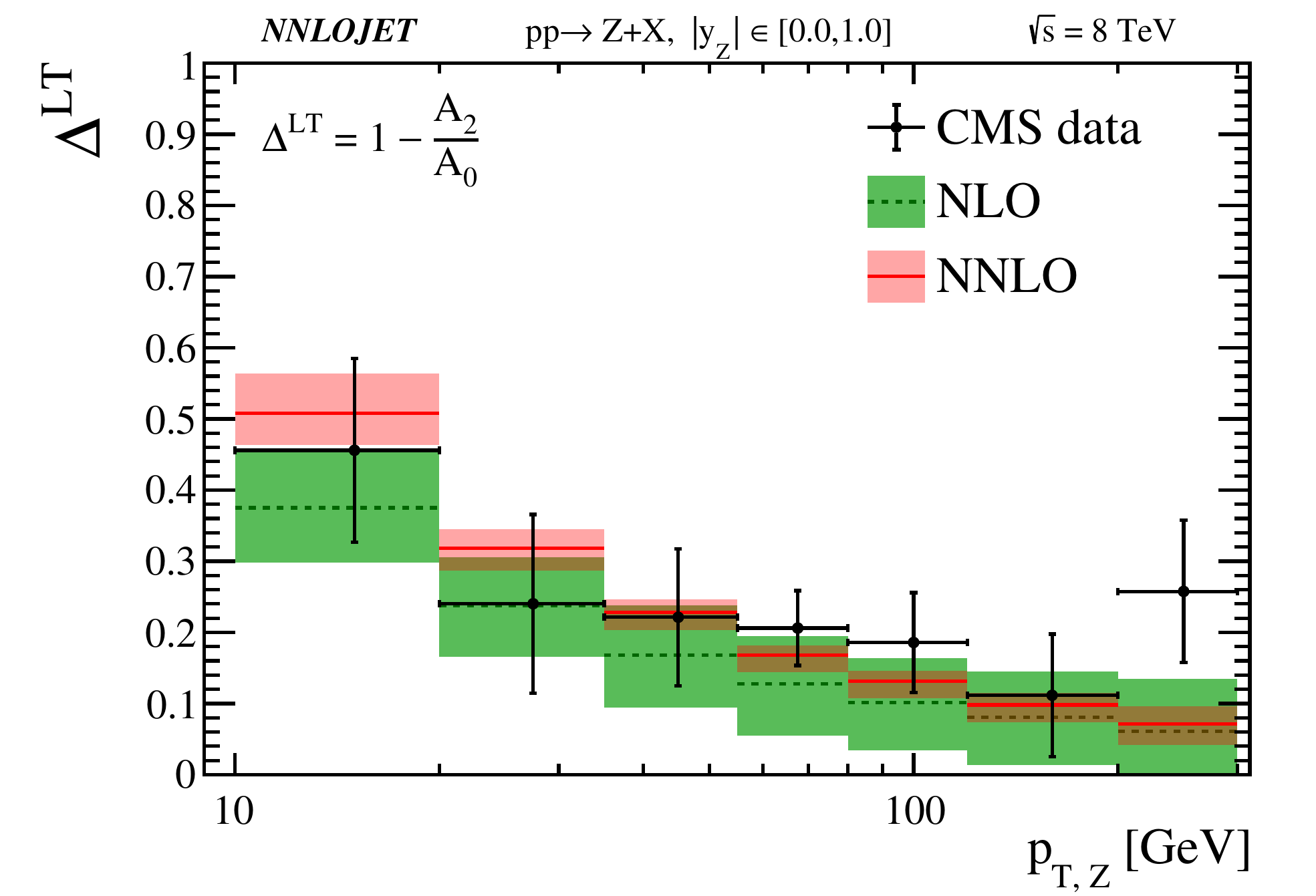} \hfill
\includegraphics[width=.49\linewidth]{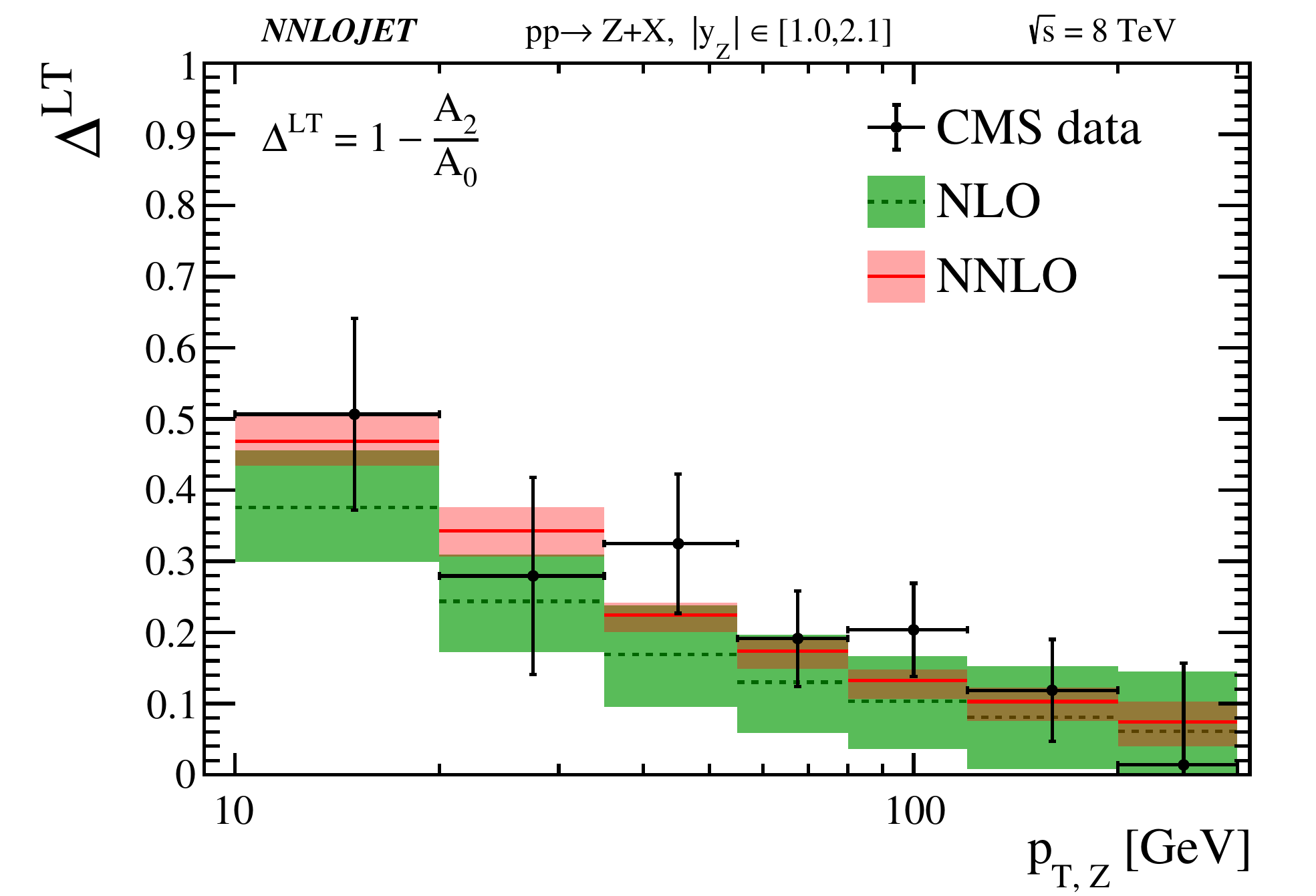}
\caption{\small The extent of the Lam--Tung violation as expressed through $\Delta^{\rm LT}$
for the CMS data in $\Pp\Pp$ collisions at $\sqrt{s} = 8~\TeV$, with the kinematic cuts of 
$\lvert\yz\rvert \in [1.0,\,2.1]$ (left) and $\lvert\yz\rvert \in [1.0,\,2.1]$ (right).
The data is compared to the corresponding NLO (green fill) and NNLO (red fill) predictions.
}
\label{fig:LamTungCMS}
\end{figure}

As discussed previously, it is also informative to express the data in terms of the new obserable $\Delta^{\rm LT}$ as 
defined in Eq.~\eqref{eq:dLT}. This comparison is performed in Figs.~\ref{fig:LamTungATLAS} and \ref{fig:LamTungCMS}
for the ATLAS and CMS measurements, respectively, where the data has been re-expressed in terms of this quantity.%
\footnote{
	We omit the lower panels with the $K$-factors in these figures, as they are almost identical to the case of $(A_0-A_2)$ shown in Figs.~\ref{fig:ATLAS_A0mA2} and \ref{fig:CMS_A0mA2} due to the small corrections to the $A_0$ coefficient.
}
It is found that the extent of the Lam--Tung violation observed in data is consistently described
by the NNLO predictions. While there is some tendency for the data to prefer a stronger Lam--Tung 
violation for $\ptz > 40~\GeV$, more precise data is required to confirm this behaviour.


\section{Conclusions and outlook} 
\label{sec:4}

Using our calculation of the $\PZ+\jet$ process at NNLO~\cite{Ridder:2015dxa}, we have computed the $\ptz$ distributions 
for the angular coefficients in $\PZ$-boson production to $\cO(\alphas^3)$.
We have focussed on the phenomenologically most relevant angular coefficients $A_{i=0,\ldots,4}$ for 
$\Pp\Pp$ collisions at $\sqrt{s} = 8~\TeV$ and have compared them with available LHC data.
With the theory uncertainties estimated by the uncorrelated variation of the factorisation and renormalisation scales (as described at the beginning of Section~\ref{sec3}), we find that these coefficients display a good perturbative convergence. In particular, a reduction of scale uncertainties is observed at each successive order and the residual scale uncertainties at NNLO are typically at the level of 5\%.
The NNLO corrections are observed to have an important impact on the predicted shapes of the distributions for $A_0$, $A_1$, and $A_2$. Of particular note is that the corrections to the $A_2$ distribution are both large and negative (up to $-20\%$) in the direction of data for both ATLAS and CMS measurements. 
It is found that the impact of the NNLO corrections to $A_3$ and $A_4$ distributions is small, and that these distributions are well described by the central NLO prediction.
Besides comparing our predictions to the available LHC data for these coefficients, we have also provided predictions for the $\ptz$ distributions for $A_{i=0,1,2}$ within the LHCb fiducial region, which would allow to probe the $\PZ$-boson production mechanism at forward rapidities. 
We find that the corrections to these distributions exhibit a similar behaviour to that at central rapidities both in size and shape.

Particular emphasis has been placed also on testing the consistency between the Lam--Tung violation 
observed in CMS and ATLAS data with respect to the theory predictions. 
To this end, we have studied the $\ptz$ distributions for the observable $(A_0-A_2)$ directly, where
the quality of the data--theory comparison has been assessed through a $\chi^2$ test.
Here, the inclusion of NNLO corrections leads to a significant improvement in the $\chi^2/N_\mathrm{dat.}$ values:
In the case of ATLAS, the $\chi^2/N_\mathrm{dat.}$ value reduces from 4.89 at NLO to 1.80 at NNLO; whereas for CMS, 
a reduction from 1.75 at NLO to 1.01 at NNLO is observed. With respect to the NNLO predictions,
no significant deviation is observed for the ATLAS data, and the CMS data is found to be fully consistent.

We further introduced a new observable  $\Delta^{\rm LT}$ defined in Eq.~\eqref{eq:dLT}, 
which is designed to better expose the violation of the Lam--Tung relation in the lower $\ptz$ regime. 
Expressed through this quantity, it becomes clear that the extent of Lam--Tung violation observed within the
range of $\ptz \in [10,40]~\GeV$, where this effect is the strongest, is consistent with the NNLO predictions. 
There however still remains some tendency for the data to systematically exceed the corresponding predictions 
at larger \ptz values. More precise data is required to clarify this situation.

Throughout this work, we have shown how the NNLO QCD predictions obtained 
via the calculation of the $\PZ+\jet$ process at $\cO(\alphas^3)$ are essential
to provide an adequate description of the \ptz distributions of several angular coefficients 
present in $\PZ$-boson production. 
It is therefore likely that a similar statement will also apply to
the case of $\PW$-boson production.
At present, a precise extraction of $M_{\PW}$ at the LHC relies on an accurate modelling of the corresponding 
angular coefficients in $\PW$-boson production~\cite{Aaboud:2017svj} based
on the fixed-order $\cO(\alphas^2)$ prediction. 
Our studies indicate that this level of theoretical accuracy is inadequate. 
The $\cO(\alphas^3)$ corrections to the decay lepton distributions in vector-boson production computed here are providing 
an important step towards improving the theoretical description of reference quantities necessary for the precise measurement of $M_{\PW}$.

\acknowledgments
  
The authors thank Daniel Froidevaux,  Elzbieta Richter--Was and Massimiliano Grazzini for many useful discussions and also Aaron Armbruster for providing us with the relevant covariance matrix of the ATLAS measurement.
We further thank Xuan Chen, Juan Cruz-Martinez, James Currie, 
Tom Morgan, Jan Niehues, and Joao Pires for useful discussions
and their many contributions to the \textsc{NNLOjet} code.
We acknowledge the computing resources provided to us by the Swiss National Supercomputing Centre (CSCS) under the project ID p501b.
This research was supported in part by the UK Science and Technology Facilities Council, by the Swiss National Science Foundation (SNF) under contracts 200020-162487 and CRSII2-160814, by the Research Executive Agency (REA) of the European Union under the Grant Agreement PITN-GA-2012-316704 (``HiggsTools'') and the ERC Advanced Grant MC@NNLO (340983).


\bibliography{Rhorry}

\end{document}